\documentclass[useAMS,usenatbib]{mn2e}
\usepackage[dvips]{epsfig}
\usepackage{color}
\usepackage{float}

\begin{document}

\label{firstpage}

\title[Modeling Hercules]{Life and death of a hero -- Lessons learned
  from modeling the dwarf spheroidal Hercules: an incorrect orbit?}  

\author[Bla\~{n}a et al.]{
  M. Bla\~na$^{1}$ \thanks{mblana, mfellhauer, rsmith, rcohen,
    jfarias, gcandlish @astro-udec.cl}, 
  M. Fellhauer$^{1}$,
  R. Smith $^{1}$,
  G.N. Candlish$^{1}$,
  R. Cohen $^{1}$,
  J.P. Farias $^{1}$ \\
  $^{1}$ Departamento de Astronom\'{i}a, Universidad de
  Concepci\'{o}n, Casilla 160-C, Concepci\'{o}n, Chile}

\pagerange{\pageref{firstpage}--\pageref{lastpage}} \pubyear{2013}

\maketitle

\begin{abstract}
  Hercules is a dwarf spheroidal satellite of the
  Milky Way, found at a distance of $\approx 138$~kpc, and showing
  evidence of tidal disruption.  It is very elongated and exhibits a
  velocity gradient of $16 \pm 3$~km\,s$^{-1}$\,kpc$^{-1}$.  Using
  this data a possible orbit of Hercules has previously been deduced
  in the literature.  In this study we make use of a novel approach to
  find a best fit model that follows the published orbit.  Instead of
  using trial and error, we use a systematic approach in order to find
  a model that fits multiple observables simultaneously.  As such, we
  investigate a much wider parameter range of initial conditions and
  ensure we have found the best match possible.  Using a dark matter
  free progenitor that undergoes tidal disruption, our best-fit model
  can simultaneously match the observed luminosity, central surface
  brightness, effective radius, velocity dispersion, and velocity
  gradient of Hercules.  However, we find it is impossible to
  reproduce the observed elongation and the position angle of Hercules
  at the same time in our models.  This failure persists even when we
  vary the duration of the simulation significantly, and consider a
  more cuspy density distribution for the progenitor.  We discuss how
  this suggests that the published orbit of Hercules is very likely to
  be incorrect.   
\end{abstract}


\section{Introduction}
\label{sec:intro}

The Milky Way (MW) has many faint galaxies as satellites.  This
population of galaxies is constituted by a rich variety, such as
dwarf spheroidals, a dwarf irregular (Small Magellanic Cloud, SMC),
and a dwarf disc galaxy   
(Large Magellanic Cloud, LMC) \citep[for an overview see
e.g.][]{Mcc12}.  The dwarf spheroidal 
galaxies (dSph) are very faint \citep[e.g.][]{mat98}, show irregular
morphologies \citep[e.g.][]{irw95} and velocity dispersions far too
high to be explainable by their luminous matter alone
\citep[e.g.][]{wal09}.  Assuming virial equilibrium makes them the
most dark matter (DM) dominated objects in the Universe.  This
fact is in concordance with the $\Lambda$CDM paradigm \citep[e.g.\
Millennium~II simulation of][]{boy09} that a galaxy like the MW should
be surrounded by many hundreds of smaller DM haloes \citep[Via Lactea
INCITE simulation of][]{kuh08}, which possibly host luminous
components, i.e.\ the dSph galaxies. 

With the advent of large surveys like the Sloan Digital Sky Survey
\citep[SDSS][]{yor00} many more of these faint and also ultra-faint
dSph galaxies are being discovered \citep[e.g.][and many more]{bel07}.
Again these galaxies show high velocity dispersions
\citep[e.g.][]{sim07} and some of them show signs of a possible
tidal elongation \citep[e.g.][]{dea12}. 

If these elongations are in fact tidal tails then these objects
cannot be currently DM dominated.  They either have formed without DM
as tidal dwarf galaxies \citep[TDG; ][for the disc of satellites
theory]{met07} or they have lost a great part of their DM halo through
tidal stripping while orbiting 
the MW \citep[e.g.][for the general picture]{smi12}.  The only
way to lose mass from the luminous component without affecting the
DM halo dramatically would be resonant stripping seen in interacting
disc galaxies \citep[see][for a study of this mechanism in the case
of dwarf disc galaxies]{deon09} or resonant tidal disruption when a
dwarf disc orbits a larger galaxy like the MW \citep{may07}.

The Hercules dSph galaxy was discovered recently by \citet{bel07}.
Its central surface-brightness is $\mu_0 \approx
27$~mag\,arcsec$^{-2}$.  Hercules lies at a distance of $\approx
138$~kpc from the Milky Way \citep{ade09a, san09}, and its luminosity
is likely between $2.68 \times 10^{4}$~L$_{\odot}$ \citep{san09} and
$3.87 \times 10^{4}$~L$_{\odot}$ \citep{col07}. Taking the mean
from these two values and using a generic mass-to-light (M/L) ratio
of $1.0$ amounts to a stellar mass of $\approx 3.3 \times
10^{4}$~M$_{\odot}$.  

Hercules contains no gas and presents no recent star formation
\citep{ade11}.  These authors also study the chemical abundances of
[Fe/H], [Ca/H] and a trend in the [Ca/Fe] abundance, which suggests an
early rapid chemical enrichment through  type~II supernovae,
followed by a phase of slow star formation dominated by enrichment
through type~Ia supernovae.  A comparison with the isochrones
indicates that the red giants in Hercules are older than $10$~Gyr,
which could give us some hints about the age of this object.  

The elongated structure observed in Hercules by \citet{col07} using
the Large Binocular Telescope (LBT) suggests that it may be in the
process of tidal disruption.  \citet{ade09a,ade09b} determine a
line-of-sight velocity dispersion of $\sigma_{\rm los} = 3.72 \pm
0.91$~km\,s$^{-1}$, and also a radial velocity gradient of $-16 \pm 
3$~km\,s$^{-1}$\,kpc$^{-1}$, measured in distance to the semi-minor
axis with respect to the right ascension.  From this the authors
deduce that Hercules has a component of stars showing rotation.   

This gradient could also be associated with an effect of tidal
distortion caused by the Milky Way instead of rotation.  Assuming that
Hercules is in tidal disruption, \citet{jin10} proposed an orbit.
Using the measured radial velocities of the stars of Hercules and the
orientation of the elongation of Hercules, the authors calculate a
tangential velocity using energy and angular momentum arguments.

\citet{san09} estimate a projected half-light radius 
of $r_{\rm h} = 230 \pm 30$~pc and a projected ellipticity of
$\epsilon = 0.67 \pm 0.03$.  Together with the assumptions of \citet{jin10}
a de-projected half-light radius and ellipticity of $r_{\rm h,deproj}
\approx 1.5$~kpc and $\epsilon_{\rm deproj} \approx 0.95$ are estimated.

The goal of this paper is to test if we can reproduce the observations
of the  dwarf galaxy Hercules under the assumption that the published
orbit is correct and that indeed Hercules is undergoing tidal
disruption.  In the next section we explain the setup of our
simulations followed by the sections describing our results.  We end
this paper with some conclusions and a discussion of our results. 

\section{Setup}
\label{sec:setup}

Dwarf spheroidals are thought to be the oldest galaxies in the
universe, forming around re-ionisation \citep{kop09}.  In isolation
they are not expected to strongly change their 
structural parameters.  This changes when they start to orbit in the
tidal field of the MW.  As the infall time of Hercules is unknown we
will initially fix the simulation time to be $10$~Gyr.  Although this
choice is rather arbitrary it mimicks the fact that the stars of
Hercules are old.  However, we also consider a 5~Gyr simulation time
later in the paper.    

We choose to start our simulations with a `DM free',
    one-component spherical object - specifically we choose a Plummer
    sphere. As a result its properties are fully described by only two
    parameters, e.g total mass and scalelength. This reduces
    significantly the parameter space we would need to study if we
    were to include a model with a dark matter halo.  Our models often
    become extended along their orbital trajectory by tidal
    disruption, resulting in final models that are elliptical in
    shape.  In fact, the process of being tidally extended was implicit
    in the derivation of the orbit of Hercules that we will assume
    throughout this paper.  However, as we will demonstrate none of our
    models can match the observed ellipticity of
    Hercules.  \citet{smi12,pen08} demonstrate that a dwarf galaxy's
    dark matter halo provides an effective shield, protecting the
    baryons from tidal stripping until the dark matter halo has been
    almost entirely removed.  As the ellipticity is produced by tidal
    disruption, the inclusion of a dark matter halo can only reduce
    the ellipticity further, causing even greater disparacy with the
    observations (i.e.\ a dark matter free progenitor will clearly be
    most effected by tides). Thus our main result - that the failure
    to match the observed ellipticity suggests that the orbit is not
    correct -  is not reliant on our choice of a dark matter free
    progenitor.  In fact we will also demonstrate that this conclusion
    is robust - independent on the choice of spherical distribution,
    and furthermore independent of the assumed infall time as well.

We use the published position of Hercules of RA (J2000) $16^{\rm
    h} 31^{\rm m} 02^{\rm s} \pm 14''$ and Dec (J2000) $12^{\rm o}
  47' 13.83'' \pm 5''$ on the sky as well as the mean out of all
  distance estimates, i.e.\ $138 \pm 7$~kpc as our initial position.
  For the velocities we use the published radial velocity of $v_{\rm
    r,GSR} = 144.7 \pm 1.2$~km\,s$^{-1}$.  \citet{jin10} used the
measured velocity gradient and the orientation of Hercules to
determine a tangential velocity, using energy and angular momentum
arguments. They determine an angle for the tangential velocity
  of $\theta = 78 \pm 4^{\rm o}$ and $v_{\rm t} =
  -16^{+6}_{-22}$~km\,s$^{-1}$.  Following their findings we now have  
the full phase space position of Hercules today.  These values
transformed into a Cartesian coordinate system are shown in the second
column of table \ref{tab:state}. 

\begin{table}
  \centering
  \caption{Position and velocity of Hercules today and at the start of
  our simulations. The position today is calculated using the
    published position in RA and Dec and the mean of the distance
    estimates of $138$~kpc.  The velocities are determined using the
    published radial velocity and the tangential velocity determined
    by \citet{jin10}.  The position and velocity at $-10$ and $-5$~Gyr
    are calculated using the values from today and performing a
    test-particle orbit calculation inside an analytic MW potential.} 
  \label{tab:state}
  \begin{tabular}{lrrr} \hline
  Hercules & observed & calculated & calculated \\ 
  & values & values & values \\
  t [Gyr] & 0 & -10 & -5 \\ \hline
  X [kpc] & -88.81 & -92.29 & 58.18 \\
  Y [kpc] & -53.06 &  74.48 & 24.36 \\
  Z [kpc] &  82.80 & 182.13 & 98.16 \\ 
  $V_{X}$ [km\,s$^{-1}$] & -94.95 & 32.36 & -96.98 \\
  $V_{Y}$ [km\,s$^{-1}$] & -48.10 & 16.87 & -42.75 \\
  $V_{Z}$ [km\,s$^{-1}$] &  99.32 & -15.24 & -124.71 \\ \hline
  \end{tabular}
\end{table}

We now have to assume a suitable potential for the MW, which fits the
observations as well as theoretical predictions.  Again we follow
\citet{jin10} and use a superposition of several analytical potentials
to simulate the different structures of the MW.
 
We use a Miyamoto-Nagai profile \citep{mia75} for the disk, defined by
\citet{pac90}: 
\begin{eqnarray}
  \Phi^{\rm MN}_{\rm disk}\left(R,z\right) & = & \frac{-GM_{\rm disc}} 
  {\left(R^2 + \left[a+\left(z^2+b^2\right)^{1/2}\right]^{2} 
    \right)^{1/2}}
\end{eqnarray}
where $R^2 = x^2 + y^2$, with the parameters $a = 3.7$~kpc, 
$b = 0.2$~kpc and $M_{\rm disc} = 8.07 \times 10^{10}$~M$_{\odot}$.

For the bulge we use a Plummer profile \citep{plu11} also defined by
\citet{pac90}: 
\begin{eqnarray}
  \label{plu}
  \Phi^{\rm Plum}_{\rm bulge} & = & \frac{-GM_{\rm bulge}}
  {\sqrt{r^2+r_{\rm Plum}^2}}  
\end{eqnarray}
with  $r_{\rm Plum} = 0.277$~kpc and $M_{\rm bulge} = 1.12 \times
10^{10}$~M$_{\odot}$. 

For the DM halo potential, we use an adiabatically 
contracted Navarro-Frenk-White halo \citep{nav97} constrained by
\citet{xue08}: 
\begin{eqnarray}
  \Phi^{\rm NFW}_{\rm halo}\left(r\right) & = & -4 \pi G \rho_{0}
  r^3_s \left[ \frac{\ln \left( 1 + r/r_s \right)} {r} -
    \frac{1}{r_c+r_s} \right]  
\end{eqnarray}
with  $r_s = 41.67$~kpc, $r_c = r_{vir} = 275$~kpc, $M_{\rm NFW} =
\frac{4 \pi} {3} \rho_{\rm cr} \Omega_m \delta_{\rm c} r_{vir}^3 =
10^{12}$~M$_{\odot}$ and $\rho_{0} = \rho_{\rm cr} \delta_{\rm c}$.

Using this potential we calculate the position and velocity of
Hercules $10$~Gyr backwards in time using a simple test particle
integration.  The position and velocities at that time are given in
the third column of Tab.~\ref{tab:state}.  We use this position and
velocities as initial conditions for our simulations. 

As explained earlier, we do not simulate any DM component of Hercules.
Instead we choose a simple Plummer model to mimic the initial state of
Hercules.  As shown in Eq.~\ref{plu}, this model is characterised by
two parameters only: the initial mass $M_{\rm pl}$ and the initial
scale-length, the Plummer radius $R_{\rm pl}$ (we use different
symbols here to distinguish from the values for the bulge potential).

Plummer spheres with different initial parameters ($M_{\rm pl}, R_{\rm
pl}$) are inserted at the calculated position and then simulated
forward in time using the particle-mesh code {\sc Superbox}
\citep{fel00}.   

As opposed to previous similar studies \citep{fell07a,smi13} we choose
to investigate a larger parameter space of initial values to assess
the general trends and behaviour of our models in a systematic manner,
instead of searching for a best matching model by a trial and error
approach.  We use Plummer masses of $5 \times 10^{4}$, $10^{5}$, $1.5
\times 10^{5}$, $1.8 \times 10^{5}$, $1.9 \times 10^{5}$, $2 \times
10^{5}$, $5 \times 10^{5}$, and $10^{6}$~M$_{\odot}$.  The Plummer
radii we vary to be $10$, $30$, $50$, $60$, $65$, $70$, $75$, $80$,
$100$, and $150$~pc.  The values chosen are not equally spaced as
there are more data points in the parameter space region of interest.
In those regions we performed even more simulations, which are not
associated with the grid-points mentioned above.  In total
we perform a suite of 70 simulations.

 We repeat part of the experiment using a simulation time of only
$5$~Gyr.  The initial position and velocities are given in the last
column of Tab.~\ref{tab:state}.  For these simulations only a small
part of the initial parameter space is used (27 simulations).

Finally, we perform a few simulations using a cuspy
Hernquist profile for the initial satellite instead of a Plummer
sphere to look for differences in the results.  We discuss the
results of these simulations in Sect.~\ref{sec:hern}.

\section{Results}
\label{sec:res}

\subsection{Parameters of Hercules}
\label{sec:pars}

In Tab.~\ref{tab:obspar} we show the observational parameters we try
to match.  

\begin{table}
  \centering
  \caption{Observational properties of Hercules we try to fit with our
    models: $M_{\rm fin}$ is the final mass (no errors given because
    of the generic $M/L$-ratio, see main text), $\mu_0$ is the central
    surface-brightness, $r_{\rm h}$ is the projected half light
    radius,  $\sigma_{\rm los}$ is the line-of-sight velocity
    dispersion, $\Delta v_{\rm r}$ the radial velocity gradient,
    $\varepsilon$ the ellipticity and $\theta$ the position angle.}  
  \label{tab:obspar}
  \begin{tabular}{lcl}
  \hline
  Final mass & $M_{\rm fin}$ &  $3.3 \times 10^{4}$~M$_{\odot}$ \\
  Central surface-brightness & $\mu_{0}$ & $27.2 \pm
  0.6$~mag\,arcsec$^{-2}$ \\  
  Projected half-light radius & $r_{\rm h}$ & $230 \pm 30$~pc \\ 
  Velocity dispersion & $\sigma_{\rm los}$ &
  $3.72 \pm 0.91$~km\,s$^{-1}$ \\
  Velocity gradient & $\Delta v_{\rm r}$ & $-10.2 \pm
  6$~km\,s$^{-1}$\,kpc$^{-1}$ \\ 
  Ellipticity & $\varepsilon$ &	$0.67 \pm 0.03$ \\
  Position angle & $\theta$ & $-78^{\rm o} \pm ~4^{\rm o}$ \\ \hline
  \end{tabular} 
\end{table}

As we are using particles with masses, we convert the mean value of
the observational total $V$-band luminosities into a mass using  
a generic mass-to-light ratio ($M/L$) of unity which gives us a final
mass of $3.3 \times 10^{4}$~M$_{\odot}$ (mean value of the two
observationally determined luminosities).  An old population like
Hercules is more likely to have a stellar mass-to-light ratio of a
few.  So, e.g.\ assuming a $M/L = 2$ would lead to a final mass of
double the value.  

To compute the mass of our model, we cannot use the remaining bound
mass only, as it is assumed that the object (Hercules) we see is
mainly a stream of unbound particles.  So we decide to use the mass of
all particles which are located in a rectangle of the size of the
visible Hercules dwarf \citep[][their figure 2]{col07}, i.e.\ a
rectangle spanning $\pm 0.2$~degrees in right ascension and $\pm
0.1$~degrees in declination from the centre of our object.  

We use the same generic $M/L$ to convert our surface densities into
surface brightnesses to match the observational value.  A value of $2$
would make our models approximately $0.6$~mag\,arcsec$^{-2}$ fainter.
To obtain the central value, we construct pixel maps with a
resolution of $80$~pixels per degree and use the value of the brightest
pixel. At the distance of Hercules the size of one pixel is
about $30$~pc.  So we can be sure to get a good mean value for the
central surface brightness, better than relying on any radial fit,
which might not be a good description of the actual profile at
all.  We have to deal with a wide range of possible results.  While
an almost undestroyed dwarf is well fitted by a Plummer profile, we
would lose the information of the faint tails providing the
elongation.  Simulations leading to tidally more disturbed models
would be best fitted by an inner profile of the bound remnant and an
outer fitting the strong tidal tails.  Finally, totally destroyed
models won't show any strong density enhancements any longer.  As
there is no global profile which could accommodate all these
different outcomes of our simulations and the fitted central surface
brightness might be very different depending on the profile used, we
decided not to rely on a fit to determine this quantity.  

To obtain the effective radius we have no other choice than to
rely on a radial fit. We measure the surface brightness,
i.e.\ surface density, in concentric rings around the centre
of Hercules, but as seen by an observer projected on the sky and then
use the actual distance of our model object to convert degrees into
parsec.  We fit a Plummer profile (as we start out with an initial
Plummer sphere) to the data points and use the Plummer
radius from this fit as half-light radius.  It can be shown
analytically that the Plummer radius contains half of the mass in
projection.  As we do not want to measure the profile of a remaining
dense and bound core, we restrict the fitting routine to values
between $0.1$ and $0.5$~degree around the centre of Hercules to ensure
we measure the profile of the tidally disturbed envelope.  This
procedure leads to spurious results for models which are still tightly
bound and had only a small mass-loss, as will be explained in the
corresponding section below.

\citet{ade09a} use the spectra of stars within about $\pm 0.2$~degrees
in right ascension and about $\pm 0.1$~degrees in declination from the
centre of Hercules to obtain the velocity dispersion.  We use all
particles in this area to compute the total line-of-sight velocity 
dispersion of our model.

The measured velocity gradient is based mainly on two stars lying
approximately $0.35$~degrees apart from each other.  \citet{ade09a}
calculate from that a velocity gradient of about $-16$~km\,s$^{-1}$\,
kpc$^{-1}$.  But the orbit obtained by \citet{jin10} only shows a
gradient of $-10.2$~km\,s$^{-1}$\,kpc$^{-1}$.  As this is the gradient
the stars would have if they just follow the orbital path and without
any peculiar motion we regard the value of \citet{jin10} as an upper
limit, which we try to match.  To calculate the gradient we calculate
the mean radial velocity of all particles located in two thin stripes
of right ascension at $\pm 0.2$~degrees from the centre of Hercules
and give the difference. 

To calculate the ellipticity and the position angle (with respect to
the declination axis) we use again the pixel maps with $80$~pixels per
degree.  These pixel maps are then analysed using the routine {\sc
  Ellipse} of IRAF.  As position angle and ellipticity vary with
radius, we use the values at a radius of $10$~pixels (approximately
$0.2$~degrees) to match the observed quantities.  

\subsection{Technique to obtain a best match model}
\label{sec:technique}

Having measured each parameter, we then proceed with the following
routine to find the best match model:   
\begin{enumerate}
\item We plot the results as a function of initial mass $M_{\rm pl}$
  for all simulations having the same initial Plummer radius in a
  double logarithmic plot.
\item We see that the data points in the region of interest follow a
  straight line, i.e.\ follow a power-law.
\item We fit a straight line to the data points in the double
  logarithmic space to obtain the power-law.
\item We use the fitted line to calculate the value of initial
  mass (with its errors) which we need to match the observable for
  each choice of Plummer radius.
\item We follow the same procedure as (i)--(iv), now plotting the
  results as a function of the initial Plummer radius for each given
  initial mass.  Then we fit lines to the results in the area of
  interest and determine the value of the Plummer radius we would
  need to match the observable.
\item We plot all these pairs of possible matches to the observable
  in a double logarithmic plot of the initial parameters.
\item We see that those points lie on a straight line, i.e.\ follow
  a power-law, as well.
\item We fit a straight line to the data points to obtain the
  power-law.
\item Finally, we plot all the obtained power-laws and
  check if all the lines intersect in the same point (i.e. within
  the same region regarding the errors) of initial parameter
  space.  This is the location for the best fitting model. 
\end{enumerate}

\subsection{Application of the technique to Hercules}

We apply this technique in an attempt to find a best-match model for
Hercules. In order to provide a clear example, we describe the
procedure used to try to match each of the observed parameters of
Hercules individually: 

\subsubsection{Final mass}
\label{sec:mass}

\begin{figure*}
  \centering
  \epsfig{file=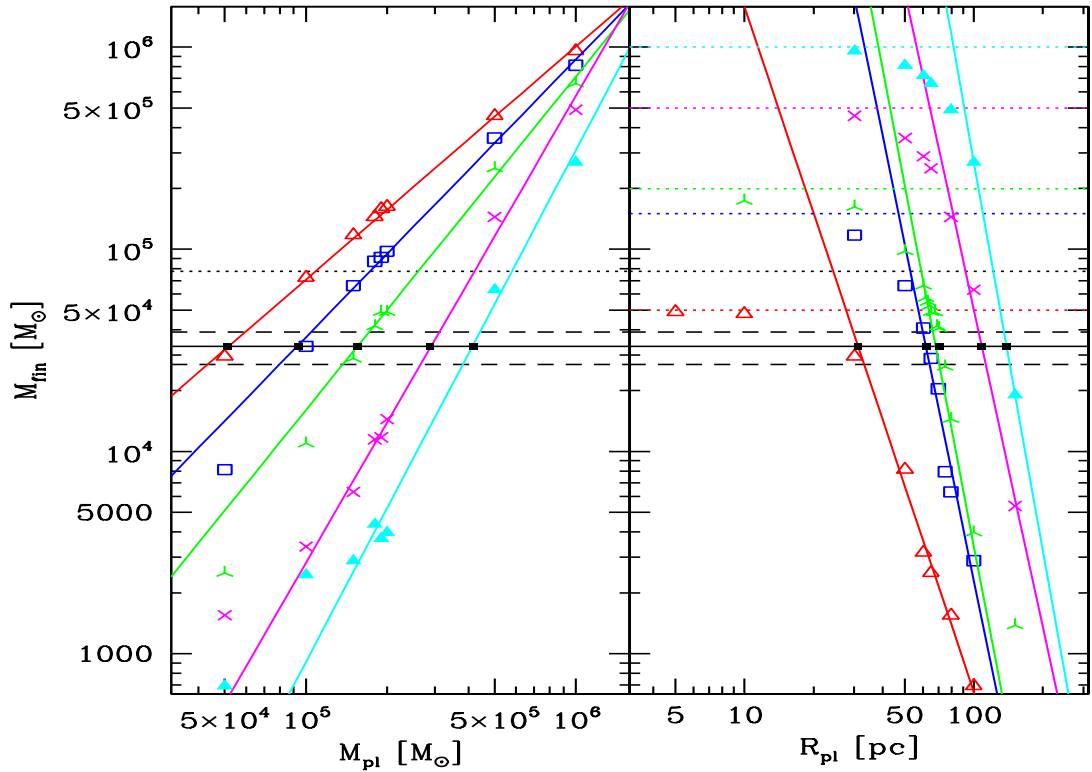, width=16cm, height=12cm}
  \caption{Left panel: Final mass of our object as function of the
    initial Plummer mass.  We show the data points and the fitting
    lines for Plummer radii of $30$ (red, open triangles), $50$ (blue,
    open squares), $65$ (green, tri-pointed stars), $80$ (magenta,
    crosses), and $100$~pc (cyan, filled triangles).  Right panel:
    Final mass as function of the initial Plummer radius.  We show
    data points and fitting lines for initial masses of $5 \times
    10^{4}$ (red, open triangles), $1.5 \times 10^{5}$ (blue, open
    squares), $2 \times 10^{5}$ (green, tri-pointed stars), $5 \times 
    10^{5}$ (magenta, crosses), and $10^{6}$~M$_{\odot}$ (cyan, filled
    triangles).  Horizontal solid line denotes the adopted value of
    the final mass we want to match, dashed lines are the
    observational errors of this value using the same $M/L$-ratio, and
    the dashed dotted line is the value we would need to match if we
    choose a $M/L$ of $2$.  Black data points (filled squares) are the
    matching values calculated by fitting power laws to the data
    points.  In the right panel additional dashed lines are shown in
    colour denoting the initial masses of the models shown in the same
    colour code.} 
  \label{fig:mass}
\end{figure*}

\begin{figure}
  \centering
  \epsfig{file=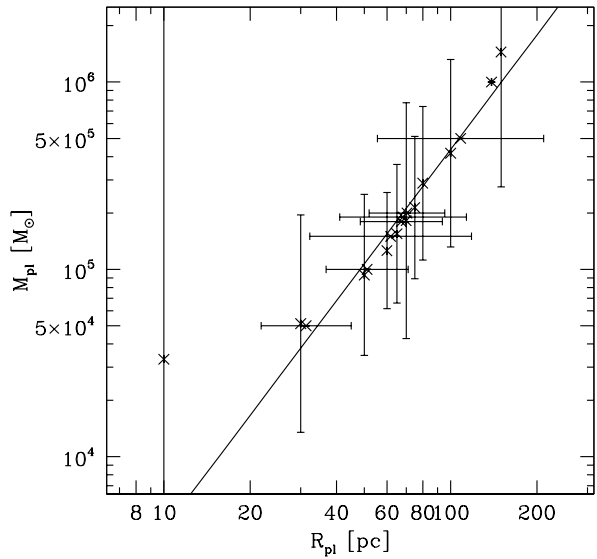, width=8.5cm, height=8.5cm, angle=0}
  \caption{Pairs of initial parameters which lead to final models with
    the correct mass of Hercules.}
  \label{fig:massfit}
\end{figure}

We measure the final mass of our objects as described in
Sec. \ref{sec:pars}.  In Fig.~\ref{fig:mass} we show for parts of our
parameter space how our resulting final masses depend on the initial
parameters.    

The left panel shows the dependency of the final mass on the initial
mass as a function of the initial scale-length.  We see that for each
choice of Plummer radius the final mass drops with smaller initial
masses, which is obvious.  Furthermore, we see that the final mass
drops more rapidly, the larger the initial Plummer radius is.  This is 
also quite obvious as more extended objects are more strongly affected
by the tides of the MW.

That the results follow strict power laws is not quite that obvious.
As long as we are not in the regime of pure tidal streams, without
a central object (lower left corner of the panel) we can easily fit
power laws of the form $\log_{10}(M_{\rm fin}[M_{\odot}]) = a *
\log_{10}(M_{\rm pl}[M_{\odot}]) + b$ to the result and are able to calculate, 
for each initial Plummer radius, which choice of initial Plummer mass
would match the required final mass.  For the curves shown, we plot
the matching values as black squares (omitting the error-bars for
clarity). 

In the right panel we plot the same data, now showing the dependency
of $R_{\rm pl}$ as function of different $M_{\rm pl}$.  Again we see
that the results follow power laws, except for when we get close to
the initial mass.  Then we have a turnover of the data points to a
much shallower dependency on the left side of each curve.  Here we
are in the very tightly bound regime (small Plummer radii) and the
objects in our simulations have not lost much of their mass.  

We again fit power laws of the form $\log_{10}(M_{\rm fin}[M_{\odot}]) = a *
  \log_{10}(R_{\rm pl}[pc]) + b$, now to the steep part of the curves
only, to obtain the matching values, which are again plotted as black
squares in the panel (again omitting the error-bars for clarity).

In Fig.~\ref{fig:mass} we only show part of the parameter space,
because otherwise the figure would be crowded.  We determine the
fitting power laws over our entire parameter space and show the
matching values, obtained in both ways, i.e.\ as function of $R_{\rm
  pl}$ and $M_{\rm pl}$ in Fig.~\ref{fig:massfit}.  In this
figure all data points have large error-bars, but only in one
direction, depending on if we determined a fitting initial mass for
a given fixed initial Plummer radius, then the error-bars are in the 
mass direction, or if we determined a fitting Plummer radius for a
given initial mass, then the error-bars are in the radius
direction.  The fitting values were determined in logarithmic
space and exhibit quite large error-bars, which are symmetric,
i.e.\ have the same length in both directions as we again plot them
in a logarithmic plot.

Despite the large error-bars the matching values seem to follow a
tight power-law.  Therefore, we again fit a simple power law of the
form 
\begin{eqnarray}
  \label{eq:power}
  \log_{10}(M_{\rm pl}[M_{\odot}]) & = & a \log_{10}(R_{\rm pl}[pc]) + b
\end{eqnarray}
through these fitted, matching values and obtain $a = 2.03 \pm 0.07$
and $b = 1.58 \pm 0.12$ which leads to
\begin{eqnarray}
  \label{eq:mfin}
  M_{\rm pl} & = 38^{+12}_{-9} R_{\rm pl}^{2.03 \pm 0.07}.
\end{eqnarray}
This line in our 2D parameter space of initial conditions gives us all
pairs of initial parameters which would fit our adopted final mass of
Hercules.  Note, that the unsymmetric error of the first value in
Eq.~\ref{eq:mfin} (and in the subsections below) solely reflects the
transformation of an error determined in logarithmic space into
regular space and not any kind of sistematics. 

\subsubsection{Surface brightness}
\label{sec:mu}

\begin{figure*}
  \epsfig{file=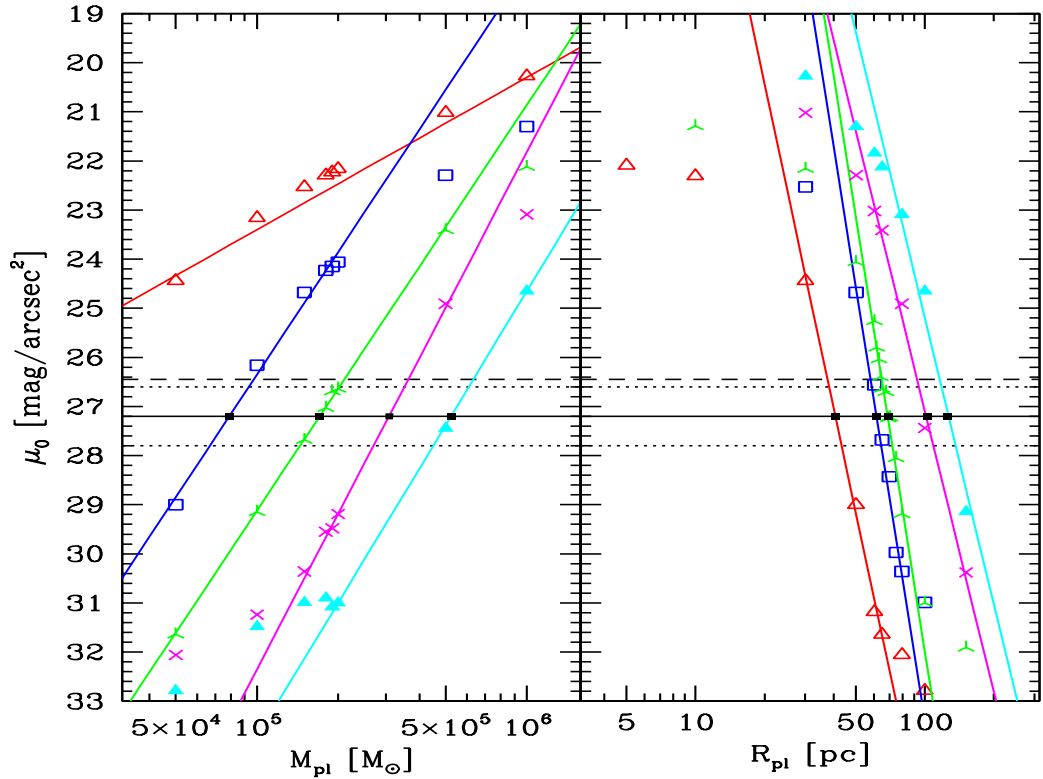, width=16cm, height=12cm}
  \centering
  \caption{Left panel: Final central surface brightness of our object
    as function of the initial Plummer mass.  We show the data points
    and the fitting lines for Plummer radii of $30$ (red, open
    triangles), $50$ (blue, open squares), $65$ (green, tri-pointed
    stars), $80$ (magenta, crosses), and $100$~pc (cyan, filled
    triangles).  Right panel: Final central surface brightness as
    function of the initial Plummer radius.  We show data points and
    fitting lines for initial masses of $5 \times 10^{4}$ (red, open
    triangles), $1.5 \times 10^{5}$ (blue, open squares), $2 \times
    10^{5}$ (green, tri-pointed stars), $5 \times 10^{5}$ (magenta,
    crosses), and $10^{6}$~M$_{\odot}$ (cyan, filled triangles).  The
    horizontal solid line denotes the observational value we want to
    match, dashed lines the observational errors and dash-dotted line
    the value we would need to match if we choose a $M/L$ of $2$, 
    i.e.\ how much the data points would shift down to lower surface
    brightnesses.  Black data points (filled squares) are the
    matching values calculated by fitting power laws to the data
    points.}  
  \label{fig:mu}
\end{figure*}

\begin{figure*}
  \centering 
  \epsfig{file=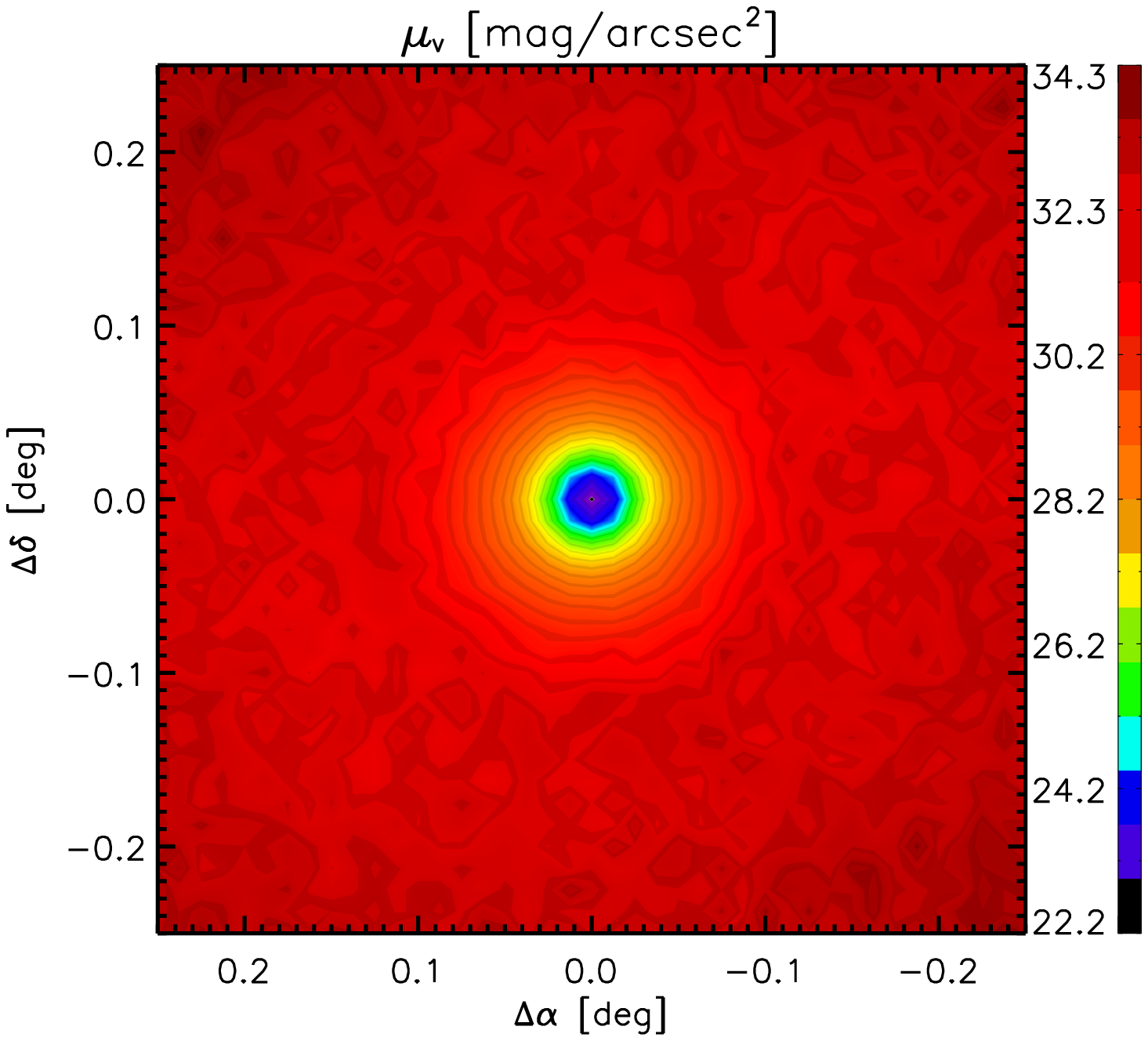, width=8.5cm, height=8.5cm}
  \epsfig{file=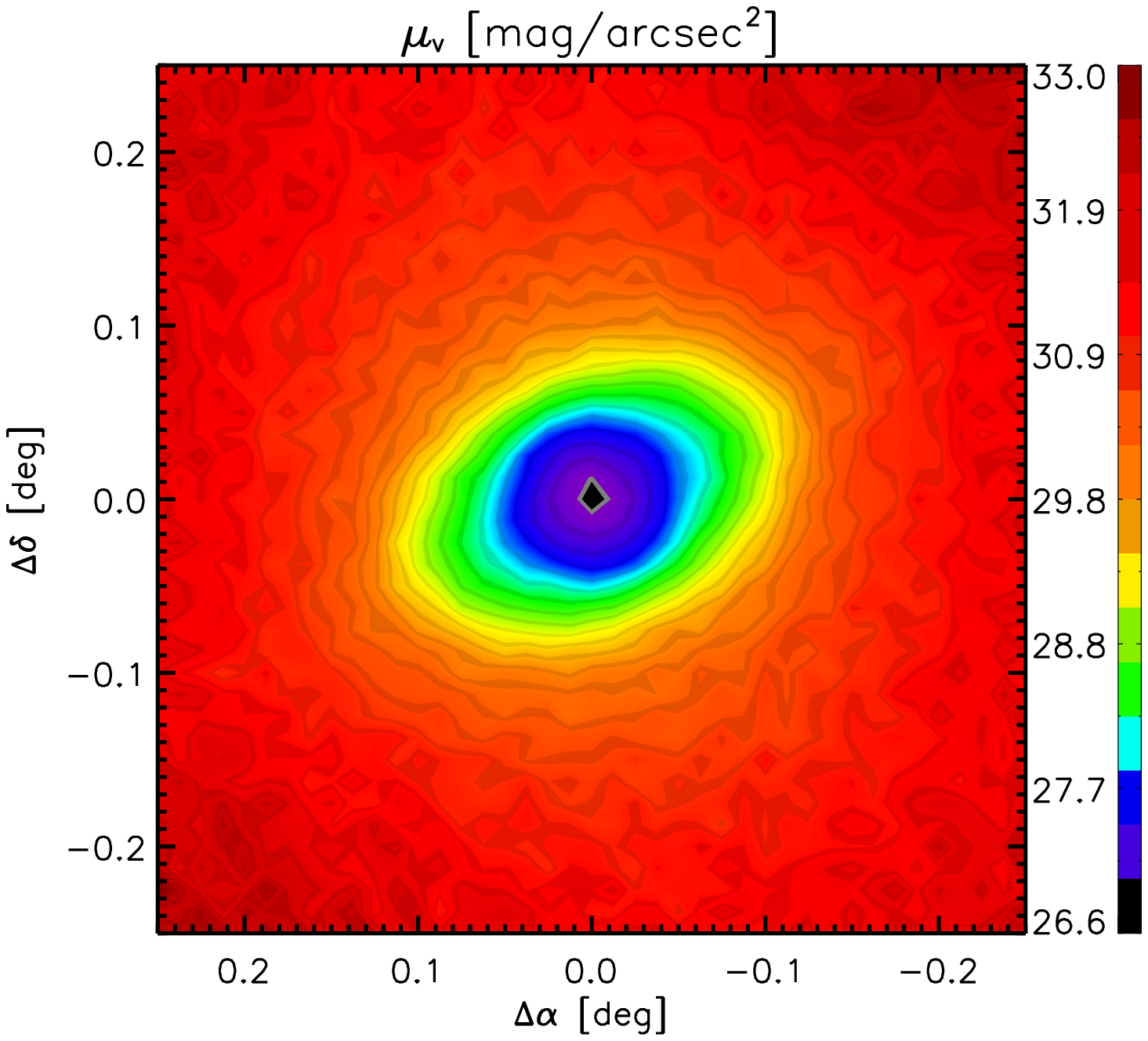, width=8.5cm, height=8.5cm}
  \epsfig{file=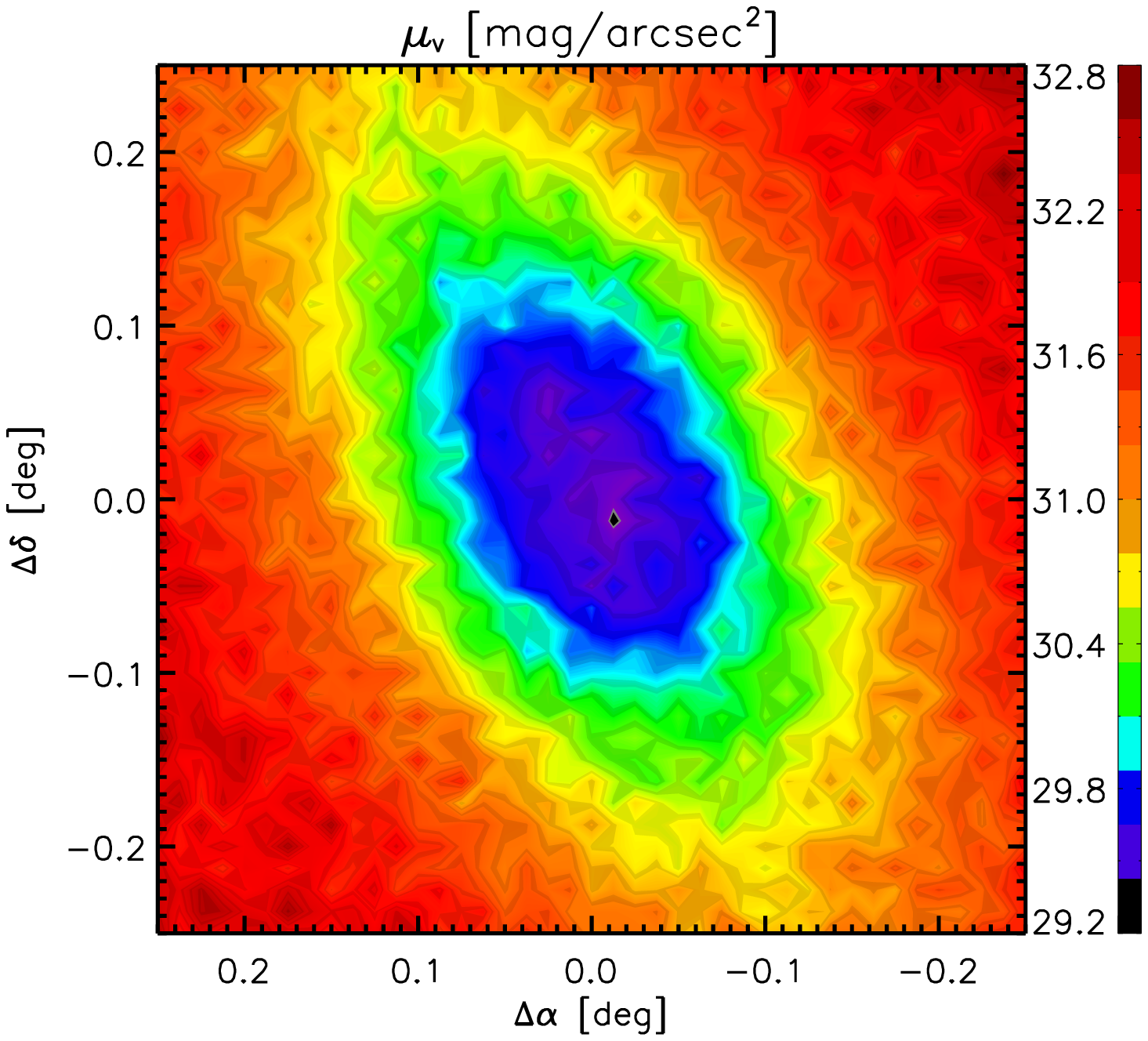, width=8.5cm, height=8.5cm}
  \epsfig{file=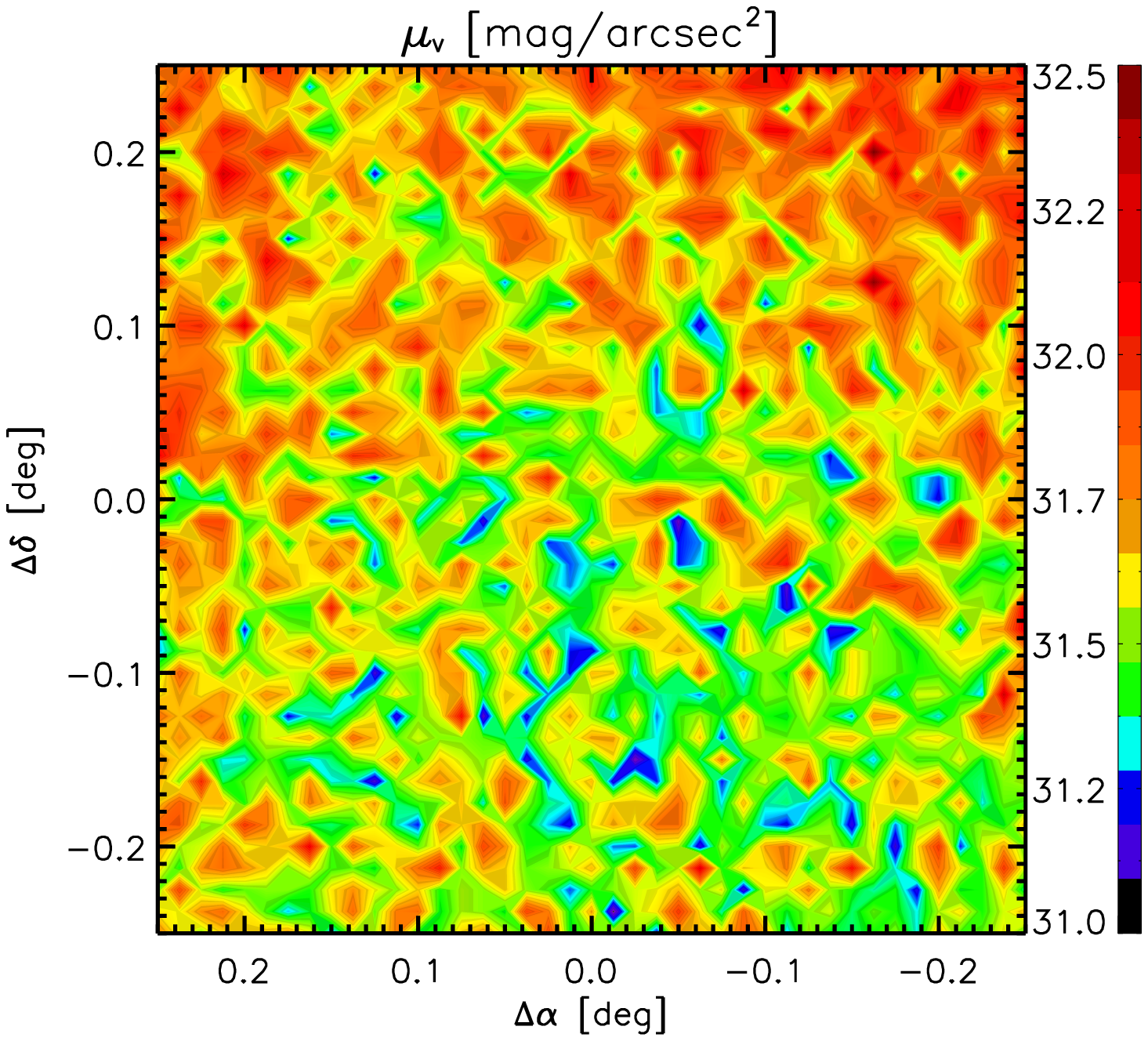, width=8.5cm, height=8.5cm}  
  \caption{Surface brightness contour plots (resolution 80 pixel per
    degree) of four of our models with an initial mass of $2.0 \times
    10^{5}$~M$_{\odot}$.  Top left panel shows the resulting object,
    using an initial Plummer radius of $30$~pc, which is in the
    `bound' regime (see main text for explanation).  The model shown
    in the top right panel has an initial Plummer radius of $65$~pc
    and is in the `tidal regime'.  The lower left panel shows the
    $80$~pc simulation in the transition phase and the lower left
    panel the $100$~pc simulation which ends up in the `stream'
    regime.  Please note the different colour maps at the side of
    each panel as we are always using the full 256-colour space to map
    the brightness differences in all panels separately.} 
  \label{fig:sdens}
\end{figure*}

\begin{figure*}
  \centering
  \epsfig{file=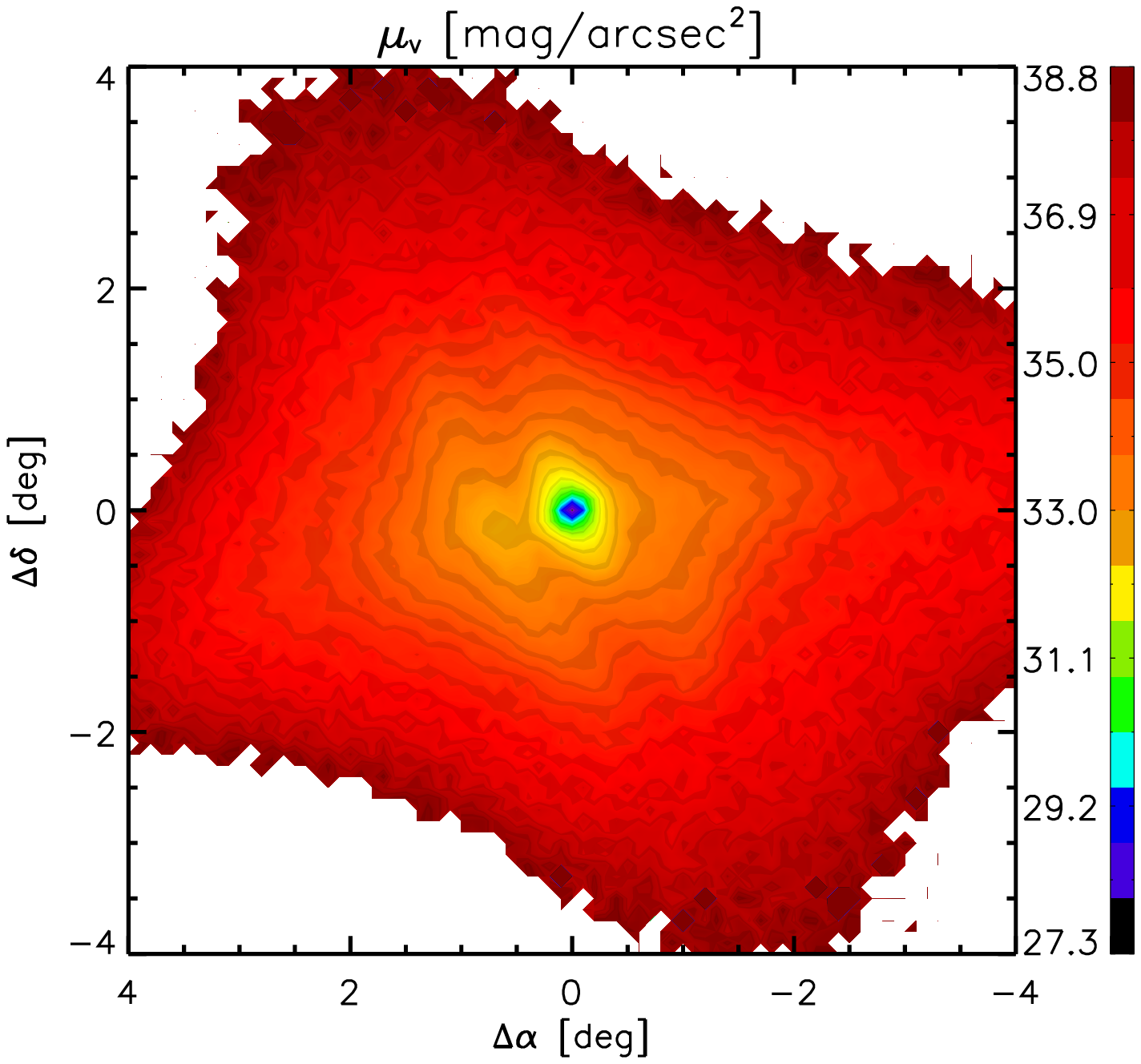, width=8.5cm, height=8.5cm}
  \epsfig{file=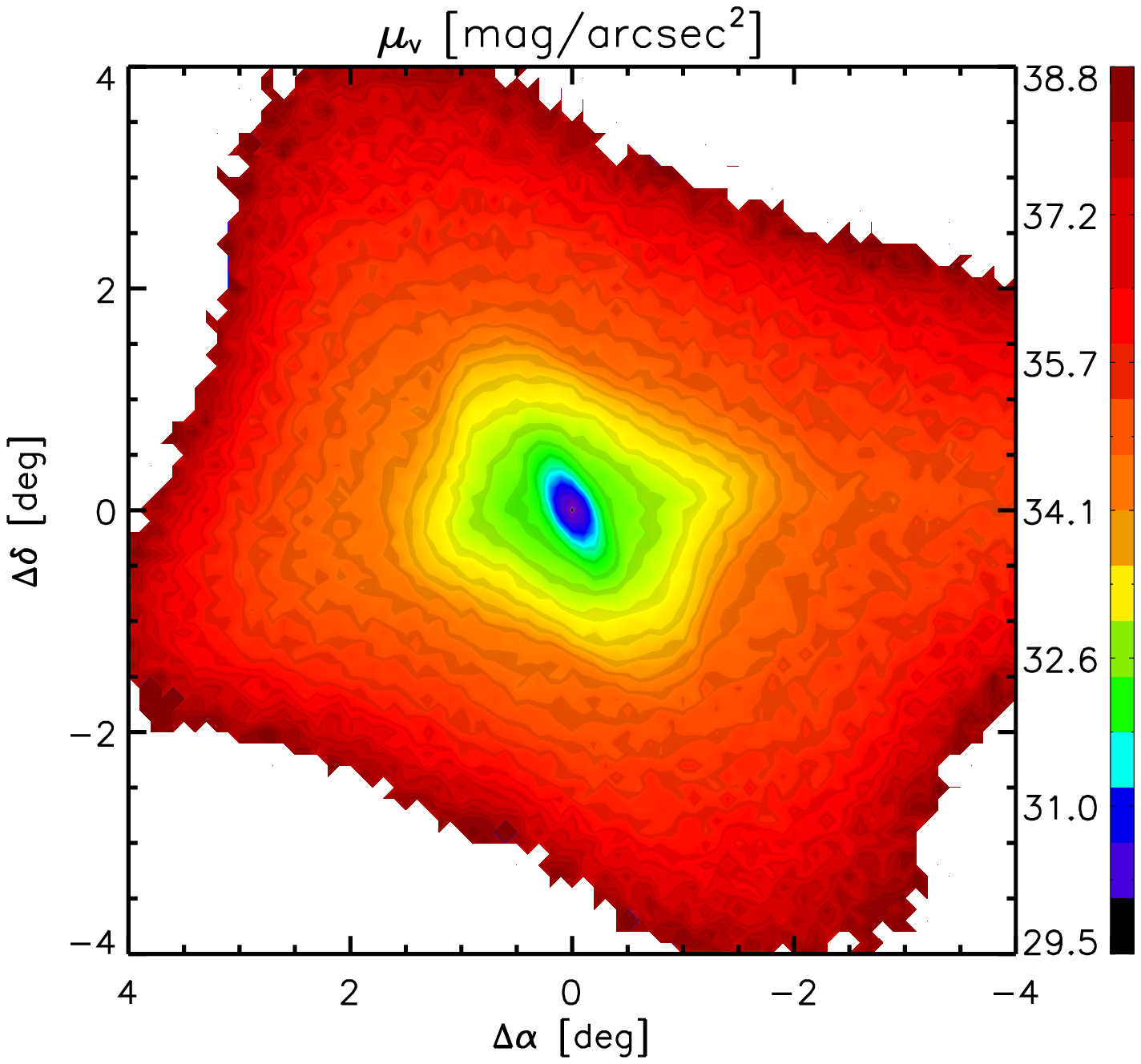, width=8.5cm, height=8.5cm}
  \caption{Surface brightness contours for two simulations shown in
    Fig.~\ref{fig:sdens} showing a larger area of the sky.  The
    strange `X-wing' shape is clearly visible in both simulations.
    Only the very innermost contours in the left panel are oriented
    along the orbit.  In both cases the intermediate contours are
    almost perpendicular to the orbit, while the outermost are
    showing the `X'-shape.  Black line in both panels denotes the
    direction to the Galactic Centre.}
  \label{fig:xwing}
\end{figure*}

\begin{figure}
  \centering
  \epsfig{file=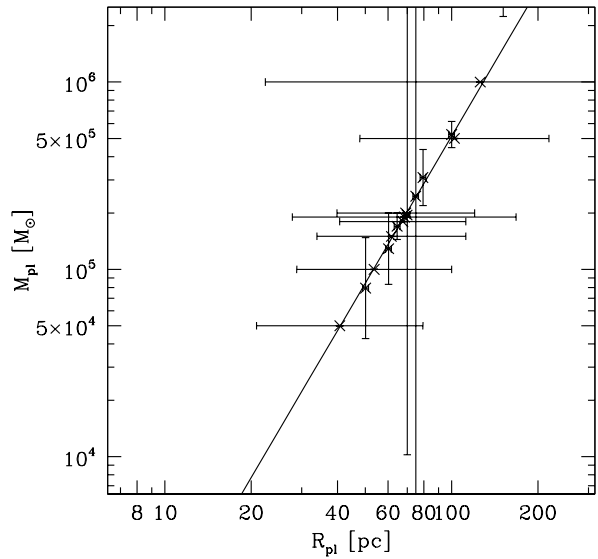, width=8.5cm, height=8.5cm, angle=0}
  \caption{Pairs of initial parameters which lead to final models with
    the correct central surface brightness of Hercules.}
  \label{fig:mufit}
\end{figure}

Now we measure the central surface brightness in all of our models and
plot the data in the same way as described in Sect.~\ref{sec:mass} in
Fig.~\ref{fig:mu}.  We use again a generic $M/L = 1.0$ to convert our
simulation surface densities (solar masses per square parsec) into
magnitudes per square arcsecond.  In the figure we also show a
dashed-dotted line which has an off-set of $0.6$~magnitudes.  This
represents the downward shift to fainter magnitudes all data points
would have if we would have used $M/L = 2.0$ instead.

In both panels of Fig.~\ref{fig:mu} we can detect three regimes of the
data points.  In the left panel we see that for higher masses the
increase in surface brightness flattens off.  Here we are in what we
refer to as the `bound' regime.  In this regime, we still have an
almost spherical central object, which is surrounded by a low density
of unbound material.  The central part of the object is almost
unaffected by the mass-loss from the outer parts.  The central surface
density (brightness) varies slowly with the strength of the mass loss
(i.e.\ the decrease in initial mass).  

Except for our simulations with Plummer radii of $30$~pc and below,
which always remain in this bound regime (see red (top) line in left
panel of Fig.~\ref{fig:mu} for the $30$~pc data), we see that the
data points turn into a steep power law dependency in the so-called
`tidal' regime.  Here our object is heavily influenced by the tidal
field and has lost half of its bound mass or more at the end of the
simulation.  The galaxy model appears to be elongated and is
surrounded by a lot of unbound material.  We use the data points in
this regime to deduce the fitting initial mass for each Plummer radius
(see left panel) and the fitting Plummer radius for each initial mass
(see right panel). 

Finally, at low masses in the left panel and for large Plummer radii
in the right panel we identify a third regime, where again the steep
power-law dependency of our results levels off.  We refer to this as
the `stream' regime.  Here we have no or almost no central bound
object any longer and the particles of our simulations simply form a
stream along the orbit of the former dSph galaxy. 

We plot surface brightness pixel maps of our simulations in
Fig.~\ref{fig:sdens} to illustrate the different regimes for
simulations with an initial mass of $2.0 \times 10^{5}$~M$_{\odot}$.
In the top left panel we see the `bound' regime (initial Plummer
radius of $30$~pc).  The object has only barely increased in size, it
is a small spherical bound object, surrounded by low density
extra-tidal material.  In the top right panel we see a typical
simulation which is in the `tidal' regime at the end of its $10$~Gyr
of evolution (initial Plummer radius of $65$~pc).  The object is much
larger than its initial size and is elongated along the adopted orbit.
It is within this regime that we search for a possible progenitor of
Hercules. 

Before we enter the `stream' regime (lower right panel, initial
Plummer radius of $100$~pc), we see a strange flip in orientation of
our object.  If we show, for example, the result of our simulation with
an initial Plummer radius of $80$~pc (lower left panel) we see an
object which is elongated almost perpendicular to its orbit.  The
reason for this strange behaviour is that we are looking at an object
which is at the brink of its destruction.  A lot of the left-over
material is now streaming through the two Lagrange points (pointing
directly towards and away from the Galactic Centre, i.e.\
perpendicular to the orbital path) into the tidal tails.  
This new material is lost during and shortly after
perigalacticon and forming `new' tidal tails, which are not yet
aligned with the orbit.  With time (i.e.\ close to apogalacticon)
they will `flip over' and align with the old tails \citep[see
e.g.][for more details]{kli09}.  In between we may see a strange
shape, which we dubbed `X-wing' tails.  Normally, the `old' tails
are denser and are responsible for the visible elongation along the
orbital path of the dwarf.  At the end of the destruction process
the dwarf loses a larger amount of mass, at say the last possible
perigalacticon before total destruction and therefore the stars in
the not-yet flipped tails might outshine the `old' tails, leading to
a flipped orientation of the elongated dwarf.  

In all cases if we look at the surface densities in a much larger
area and down to brightnesses which are not observable any longer,
we always see the `X-wing' shape formed by the unbound stars.  In
Fig.~\ref{fig:xwing} we show a larger part of the sky ($4 \times
4$~degree) for two of our simulations (same as shown in
Fig.~\ref{fig:sdens} top right and lower left panel).  We see that
in the left panel (simulation in the `tidal' regime, without flipped
contours) only the very innermost contours (which are also the only
observationally visible contours) are aligned with the orbit.
Further out the contours are `flipped'.  The very low density
contours at far distances of the remnant show the X-shape.  The
right panel shows a `flipped' simulation, i.e.\ also the innermost
contours are `flipped' (again the only contours in the observable
brightness range) as are the intermediate ones and the outer
contours show the X-shape. 

In the `stream' regime (shown in the lower right panel of
Fig.~\ref{fig:sdens}) we finally see a broad low-density stream
of material along the projected orbit.

As explained in section~\ref{sec:mass} we use the same analysis
procedure and fit power-law lines to the data points in the
`tidal'-regime.  This results in matching values of initial mass for
each choice of initial Plummer radius and in matching values of
initial Plummer radius for each value of initial mass.  Again, these
solutions have only errors in one dimension, as the other dimension
is given.  In Fig.~\ref{fig:mufit} we plot all the solutions with
their error-bars in the initial parameter space.  Again all solutions
follow a tight power-law despite the large error-bars.  If we fit a
power-law of the form of Eq.~\ref{eq:power} to the data points we
obtain $a = 2.608 \pm 0.034$ and $b = 0.495 \pm 0.063$ which
translates into the relation: 
\begin{eqnarray}
  \label{eq:mu}
  M_{\rm pl} & = & 3.1^{+0.5}_{-0.4} R_{\rm pl}^{2.608 \pm 0.034},
\end{eqnarray}
describing the one-dimensional solution space of simulations showing
the correct surface brightness at the end of the evolution.

\subsubsection{Effective radius}
\label{sec:reff}

\begin{figure*}
  \centering
  \epsfig{file=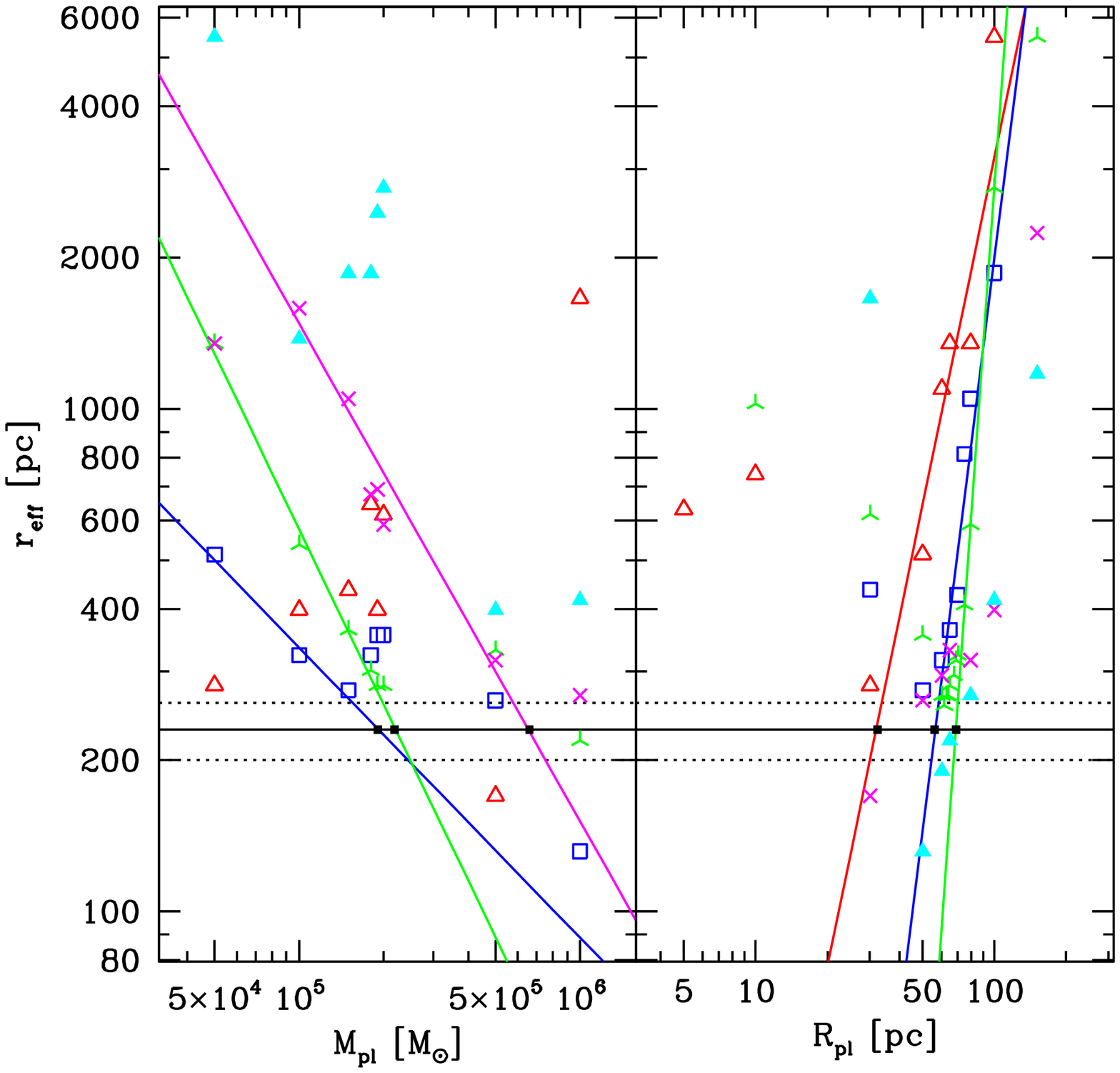, width=16cm, height=12cm}
  \caption{Left panel: Final effective radius of our object
    as function of the initial Plummer mass.  We show the data points
    and the fitting lines for Plummer radii of $30$ (red, open
    triangles), $50$ (blue, open squares), $65$ (green, tri-pointed
    stars), $80$ (magenta, crosses), and $100$~pc (cyan, filled
    triangles).  Right panel: Final effective radius as a function of
    the initial Plummer radius.  We show data points and fitting lines
    for initial masses of $5 \times 10^{4}$ (red, open triangles),
    $1.5 \times 10^{5}$ (blue, open squares), $2 \times 10^{5}$
    (green, tri-pointed stars), $5 \times 10^{5}$ (magenta, crosses),
    and $10^{6}$~M$_{\odot}$ (cyan, filled triangles).  The horizontal
    solid line denotes the observational value we want to match,
    dotted lines the observational errors.  Black data points (filled
    squares) are the matching values calculated by fitting power laws
    to the data points.} 
  \label{fig:reff}
\end{figure*}

\begin{figure}
  \centering
  \epsfig{file=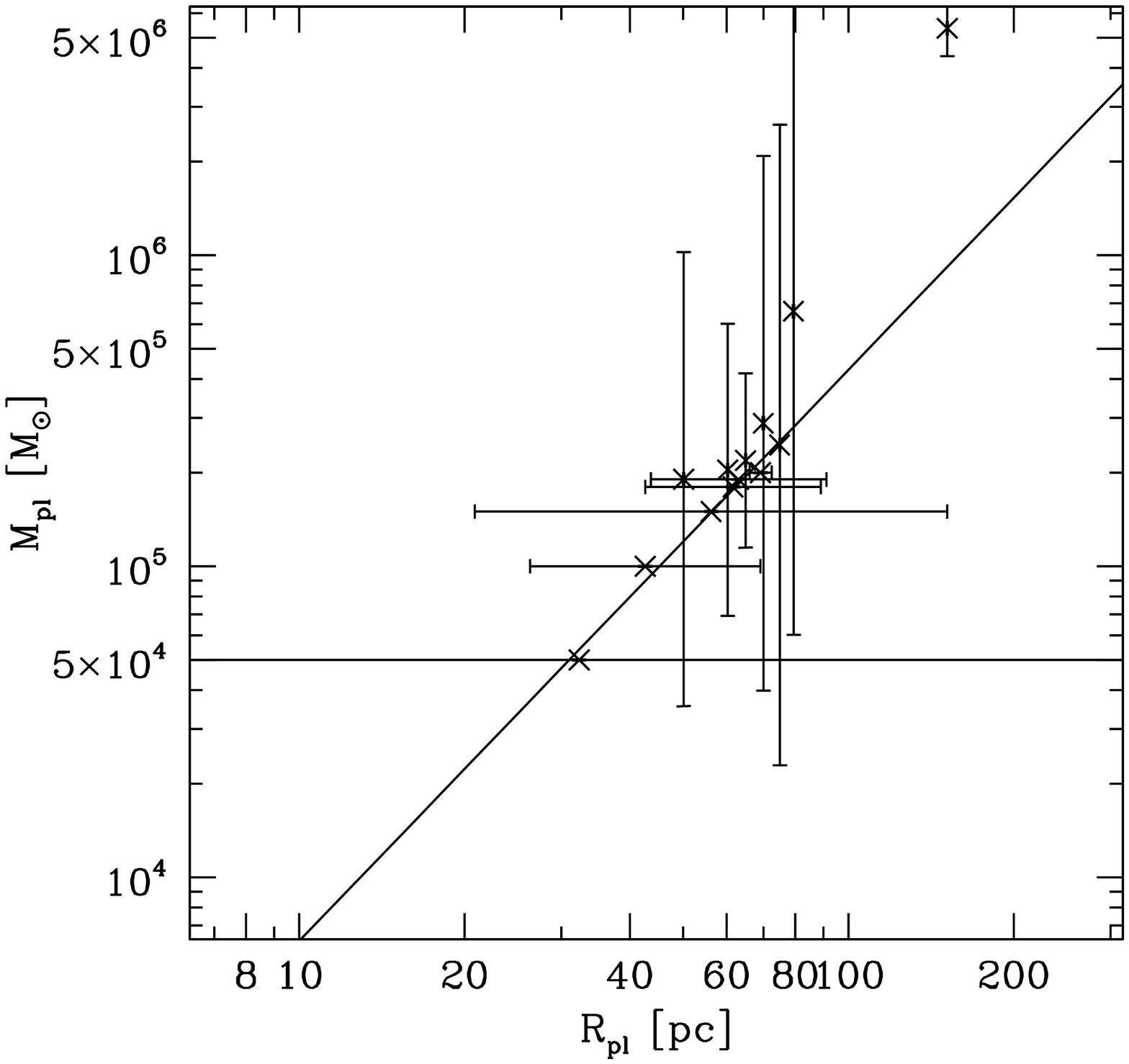, width=8.5cm, height=8.5cm}
  \caption{Pairs of initial parameters which lead to final models with
    the correct projected effective radius of Hercules.}
  \label{fig:refffit}
\end{figure}

The analysis of the results for the effective radius shown in
Fig.~\ref{fig:reff} is a bit more complicated.  As described above we
omit the very central $0.1$~degree, which might still host a bound
core, from the fitting routine.  We only fit out to $0.5$~degree as
this is the region of interest, in which the visible part (i.e.\ the
part with measurable surface brightnesses) of Hercules is located.
Furthermore, we fit the data with concentric circles and not with
ellipses.   

This procedure leads to strange results in some parts of the parameter
space.  If we are in the `bound' regime, we completely neglect the
bound object and only fit a profile to the low-density tidal
material.  The values obtained are in the regime of up to several
hundred parsecs but do not follow a strict power-law.  As a trend we
can say that as we approach the tidal regime the effective radius
becomes smaller.  This regime is shown in the right part of the
left panel and the left part of the right panel in
Fig.~\ref{fig:reff}.  This regime is followed by the `tidal' regime,
in which we again are able to fit steep power-laws to our results and
where the results reflect the effective radius of our resulting
elongated and inflated objects.  Finally, in the `stream' regime we
have almost constant density all along the stream.  The `fitted'
effective radius on the order of more than one kiloparsec reflects now
the transversal extension of the tidal stream and not a scale-radius
of any kind. 

We are not able to fit power-laws to the extreme parts of our
parameter space.  For low Plummer radii or very high initial masses
we are completely in the bound regime and for large initial radii
or very low initial masses we are completely in the `stream' regime.
For those values of initial parameters which fall into the `tidal
regime' we are able to obtain power-law fits.  We show the resulting
matching results in Fig.~\ref{fig:refffit}. 

Again the results seem to follow a power-law in initial conditions
parameter space much tighter than the error-bars suggest, but not as
tight as the final mass or the surface brightness.  The fitting values
of Eq.~\ref{eq:power} are $a = 1.837 \pm 0.080$ and $b = 1.96 \pm
0.14$, which translates to a relation of
\begin{eqnarray}
  \label{eq:reff}
  M_{\rm pl} & = & 91^{+34}_{-25} R_{\rm pl}^{1.837 \pm 0.080}.
\end{eqnarray}

\subsubsection{Velocity dispersion}
\label{sec:sigma}

\begin{figure*}
  \centering
  \epsfig{file=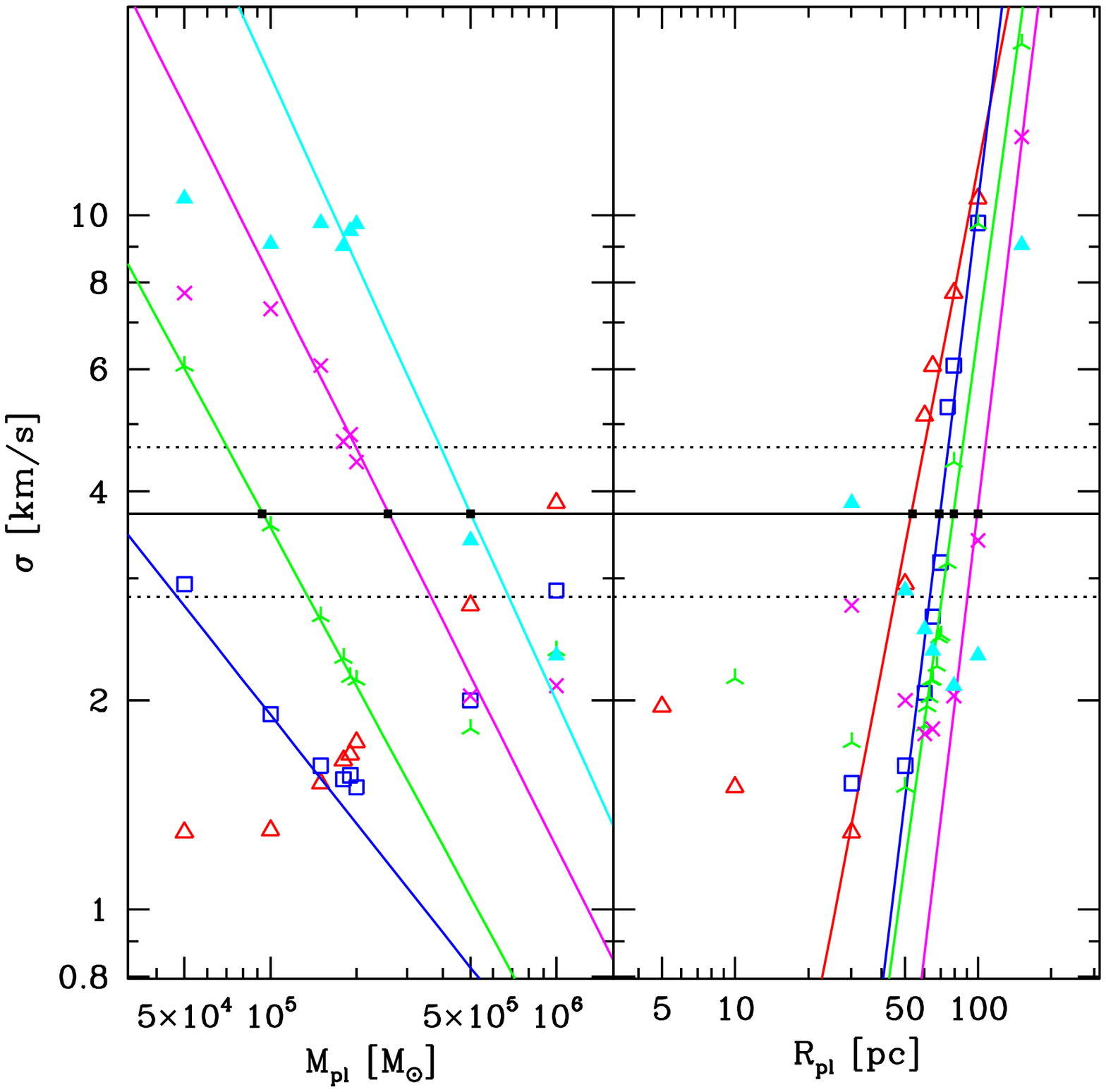, width=16.0cm, height=12.0cm}
  \caption{Left panel: Final total velocity dispersion of our object
    as a function of the initial Plummer mass.  We show the data points
    and the fitting lines for Plummer radii of $30$ (red, open
    triangles), $50$ (blue, open squares), $65$ (green, tri-pointed
    stars), $80$ (magenta, crosses), and $100$~pc (cyan, filled
    triangles).  Right panel: Final total velocity dispersion as a
    function of the initial Plummer radius.  We show data points and
    fitting lines for initial masses of $5 \times 10^{4}$ (red, open
    triangles), $1.5 \times 10^{5}$ (blue, open squares), $2 \times
    10^{5}$ (green, tri-pointed stars), $5 \times 10^{5}$ (magenta,
    crosses), and $10^{6}$~M$_{\odot}$ (cyan, filled tri-angles).  The
    horizontal solid line denotes the observational value we want to
    match, dotted lines the observational errors.  Black data points
    (filled squares) are the matching values calculated by fitting
    power laws to the data points.} 
  \label{fig:sigma}
\end{figure*}

\begin{figure}
  \centering
  \epsfig{file=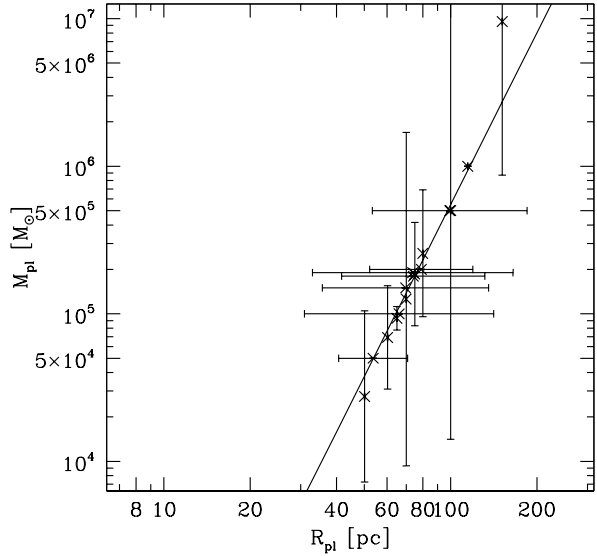, width=8.5cm, height=8.5cm, angle=0}
  \caption{Pairs of initial parameters which lead to final models with
    the correct total line-of-sight velocity dispersion of Hercules.}
  \label{fig:sigmafit}
\end{figure}

We now repeat the same procedure with the overall velocity
dispersion.  In Fig.~\ref{fig:sigma} we can clearly distinguish two
regimes in the results showing a `U'-shaped dependency.  At high
masses in the left panel and small Plummer radii in the right we are
in the `bound' regime and the velocity dispersion, even though
measured over all particles within the region described above, is
mainly due to the bound particles and is related with the bound
mass and the Plummer radius according to the virial theorem:
\begin{eqnarray}
  \label{eq:virial}
  \sigma^{2} & \sim & \frac{\rm Mass}{\rm scale\ radius}.
\end{eqnarray}
As the mass-loss in this regime is low, we still see almost the same
dependency of the final velocity dispersion on the initial values.

The second solution we find in the `tidal' regime, where the measured
velocity dispersion rises quickly with smaller initial masses (left
panel) and larger Plummer radii (right panel).  

Both regimes may lead to possible solutions but we only take the
solutions from the `tidal' regime into account. 

The third regime (`stream') is visible in the left panel with the
results for initial Plummer radii of $100$~pc.  At some point we have
a saturation of measured velocity dispersion because of the finite
extension the stream can have.

Again we fit power-laws to the results in the `tidal' regime and
calculate the matching values of initial parameters which result in
the correct velocity dispersion.  These values with their error-bars
are shown in Fig.~\ref{fig:sigmafit}. 

The fitting line shown in Fig.~\ref{fig:sigmafit} has values of $a =
3.87 \pm 0.10$ and $b = -2.00 \pm 0.19$ leading to
\begin{eqnarray}
  \label{eq:sigma}
  M_{\rm pl} & = & 0.01^{+0.02}_{-0.003} R_{\rm pl}^{3.87 \pm 0.01}.
\end{eqnarray}

\subsubsection{Velocity gradient}
\label{sec:vrad}

\begin{figure*}
  \centering
  \epsfig{file=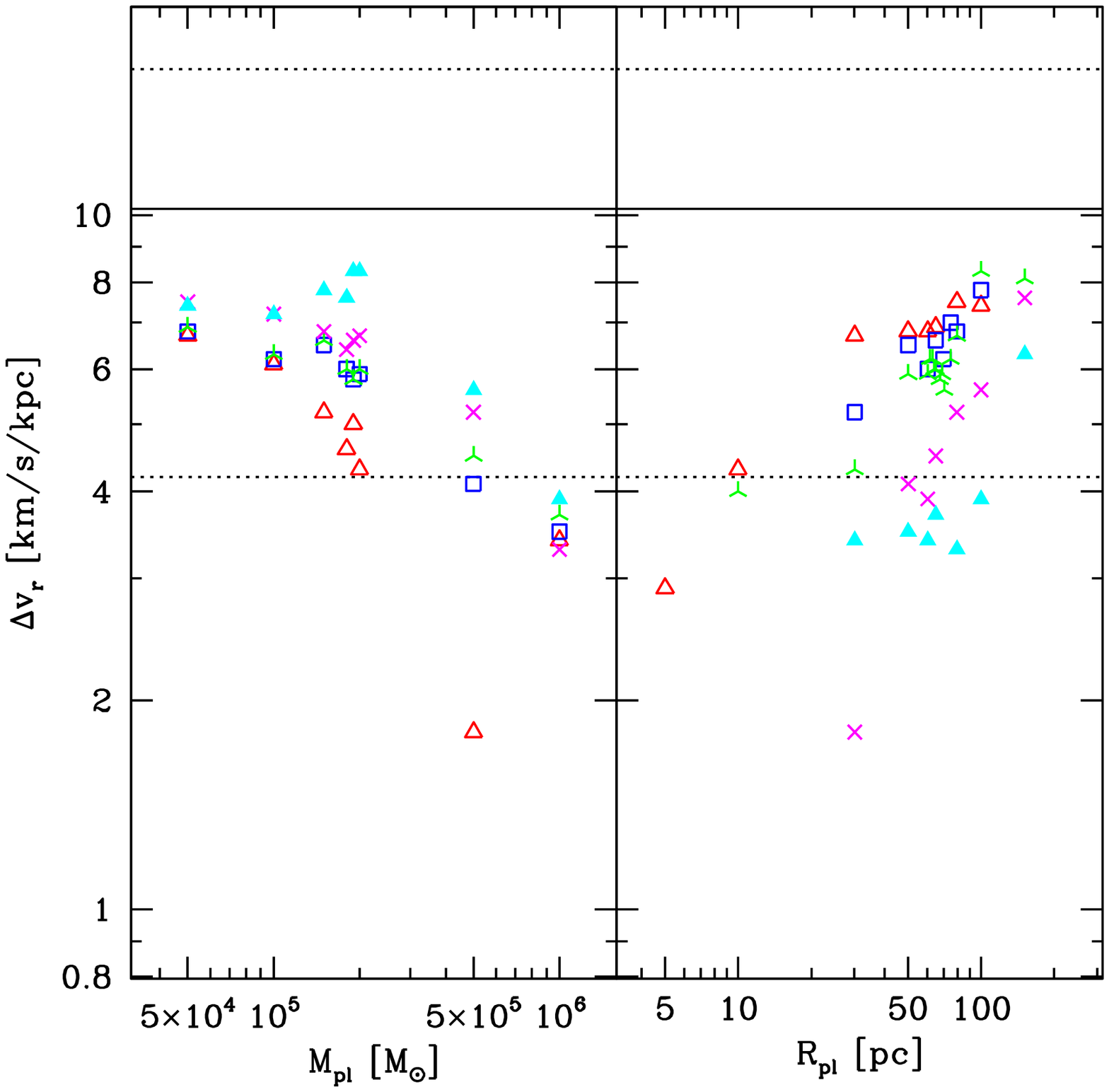, width=16.0cm, height=12.0cm}
  \caption{Left panel: Final velocity gradient of our object as a
    function of the initial Plummer mass.  We show the data points 
    for Plummer radii of $30$ (red, open triangles), $50$ (blue, open
    squares), $65$ (green, tri-pointed stars), $80$ (magenta,
    crosses), and $100$~pc (cyan, filled triangles).  Right panel:
    Final velocity gradient as a function of the initial Plummer
    radius.  We show data points for initial masses of $5 \times
    10^{4}$ (red, open triangles), $1.5 \times 10^{5}$ (blue, open
    squares), $2 \times 10^{5}$ (green, tri-pointed stars), $5 \times  
    10^{5}$ (magenta, crosses), and $10^{6}$~M$_{\odot}$ (cyan, filled
    triangles).  Horizontal solid line denotes the value we want to
    match based on the assumed orbit, dotted lines the one-sigma
    deviation of this value.}  
  \label{fig:vrad}
\end{figure*}

In Fig.~\ref{fig:vrad} we see the results of our simulations regarding
the final velocity gradient measured as described above.  We see
immediately that the results are not easily fitted with power-laws and
also it is impossible to distinguish the three regimes.  We see a
general trend to smaller gradients if we tend to the `bound' regime,
i.e.\ to larger masses and smaller Plummer radii.  None of our
simulations fits the velocity gradient adopted by \citet{jin10} but a
lot of simulations reach this value within the lower one-sigma error
shown.  The reason for our small values are two-fold: near-field tidal
tails do not need to align with the orbit completely.  They may even
be perpendicular as seen in the lower left panel of
Fig.~\ref{fig:sdens}.  So the mean radial velocities measured may not
have the exact radial velocity a particle, following exactly the
adopted orbit for Hercules, would have at this point.  Furthermore, we
are not dealing with one particle but with an extended tidal tail,
where particles are on similar but not on identical orbits and have
peculiar motions similar to epicycles as well.  So, if we calculate a
mean radial velocity at two given points, we will always get a
superposition of these two effects and values which are somewhat below
the orbital radial velocity difference.  

As a result we note that we match the adopted velocity gradient of the
orbit to within the one-sigma error for all
simulations with masses below $5.0 \times 10^{5}$~M$_{\odot}$ and for
Plummer radii larger than $50$~pc. 

\subsubsection{Ellipticity and position angle}
\label{sec:eps}

\begin{figure*}
  \centering
  \epsfig{file=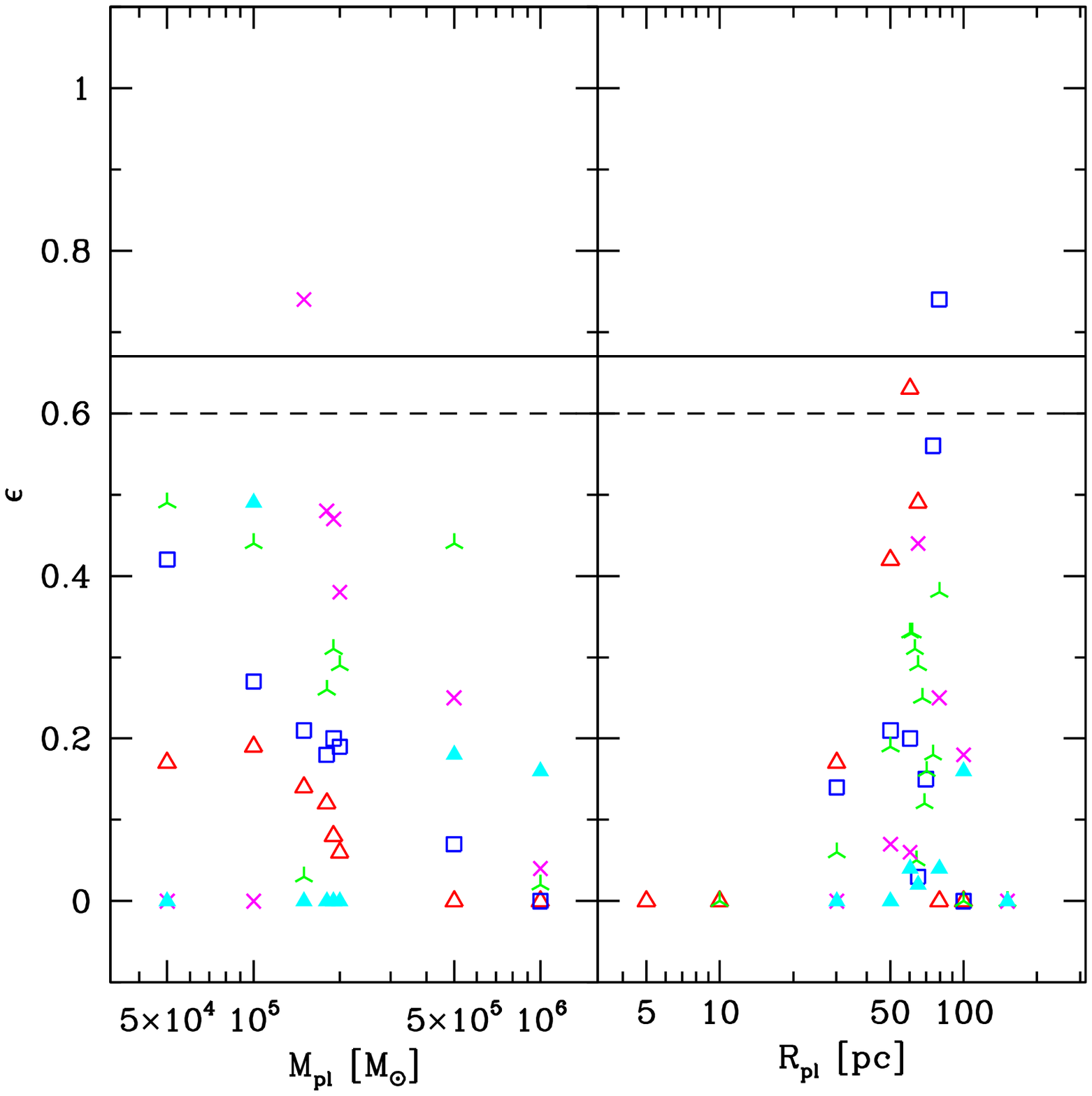, width=16.0cm, height=12.0cm}
  \caption{Left panel: Final ellipticity of our object as a function of
    the initial Plummer mass.  We show the data points for Plummer
    radii of $30$ (red, open triangles), $50$ (blue, open squares),
    $65$ (green, tri-pointed stars), $80$ (magenta, crosses), and
    $100$~pc (cyan, filled triangles).  Right panel: Ellipticity as a
    function of the initial Plummer radius.  We show data points for
    initial masses of $5 \times 10^{4}$ (red, open triangles), $1.5
    \times 10^{5}$ (blue, open squares), $2 \times 10^{5}$ (green,
    tri-pointed stars), $5 \times 10^{5}$ (magenta, crosses), and
    $10^{6}$~M$_{\odot}$ (cyan, filled triangles).  Horizontal solid
    line denotes the observational value we want to match, dotted
    lines the observational errors.}
  \label{fig:ecc}
\end{figure*}

\begin{figure*}
  \centering
  \epsfig{file=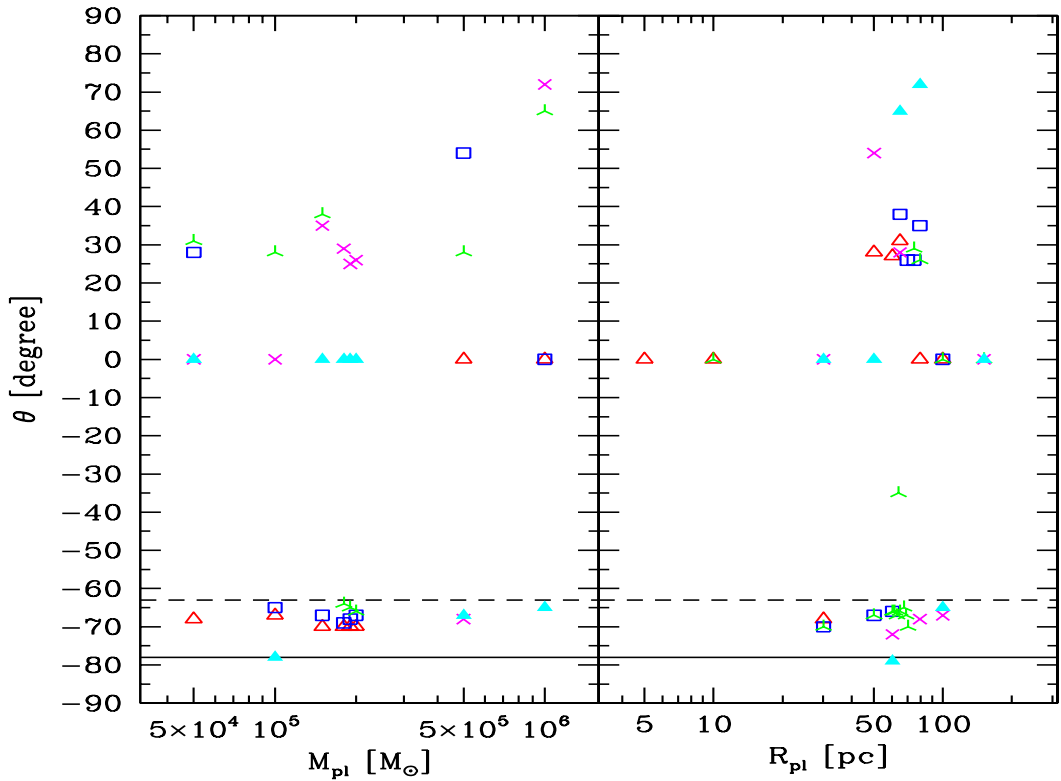, width=16.0cm, height=12.0cm}
  \caption{Left panel: Final position angle of our object as function
    of the initial Plummer mass.  We show the data points for Plummer
    radii of $30$ (red, open triangles), $50$ (blue, open squares),
    $65$ (green, tri-pointed stars), $80$ (magenta, crosses), and
    $100$~pc (cyan, filled triangles).  Right panel: Final position
    angle as a function of the initial Plummer radius.  We show
    data points for initial masses of $5 \times 10^{4}$ (red, open
    triangles), $1.5 \times 10^{5}$ (blue, open squares), $2 \times
    10^{5}$ (green, tri-pointed stars), $5 \times 10^{5}$ (magenta,
    crosses), and $10^{6}$~M$_{\odot}$ (cyan, filled triangles).
    Horizontal solid line denotes the observational value we want to
    match, dashed line the observational error.}
  \label{fig:theta}
\end{figure*}

\begin{figure}
  \centering
  \epsfig{file=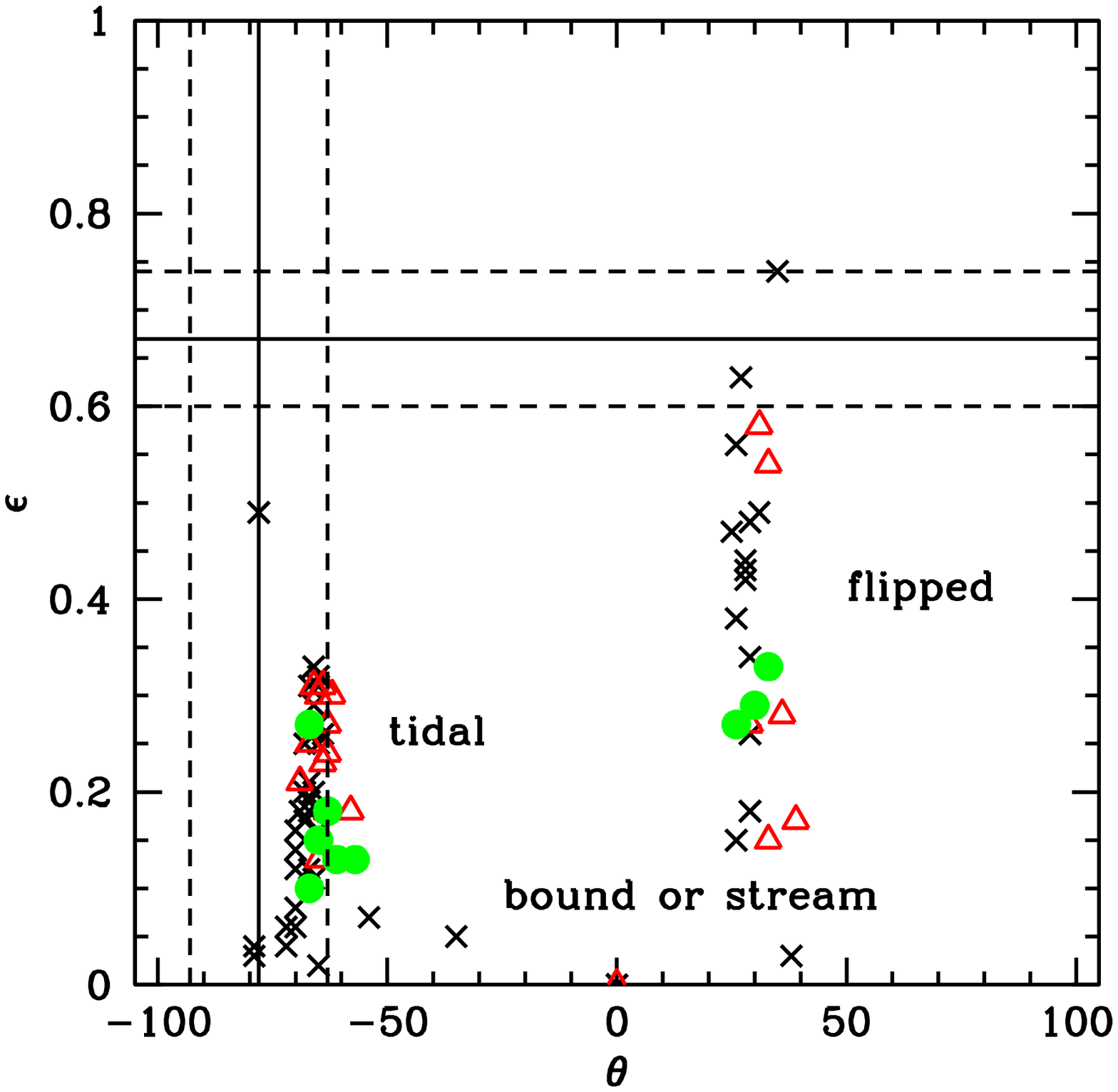, width=8.5cm, height=8.5cm, angle=0}
  \caption{ Ellipticity ($\epsilon$) vs.\ position angle ($\theta$)
      for all our simulations.  Crosses are Plummer spheres with
      $10$~Gyr of orbital time, triangles (red) are Plummer spheres
      with $5$~Gyr of orbital time and circles (green) are Hernquist
      models with $10$~Gyr of orbital time.  We clearly see that the
      results are grouped according to the different regimes
      (labels).}
  \label{fig:ellippa}
\end{figure}

In Fig~\ref{fig:ecc} we plot the ellipticity of our final object as a
function of the initial mass (left panel) and the Plummer radius
(right panel).

What we see is that none of our models can reach the correct strong
elongation of $\epsilon = 0.68$ as observed.  The only simulations
which show elongations as pronounced as the real Hercules are
simulations in the transition regime between `tidal' and `stream',
which show a perpendicular orientation of the visible contours (the
`flipped-tails' regime).  An example of this can be seen in the lower
right panel of Fig.~\ref{fig:sdens}.  We can clearly see the effect of
the `flipping tails' (`X-wing'-effect), which alter the shape of our
final object.  Therefore in order to reach the observed values of
ellipticity, we find we must forfeit the position angle. 

Fig.~\ref{fig:sdens} visually demonstrates why this is the case.
First we are in the `bound' regime and our object is still spherical,
i.e.\ in this regime the position angle is not defined and set equal
to zero.  Then we are entering the `tidal' regime, where the position
angle is similar to the observed value.  This changes dramatically when
we enter the transition regime between `tidal' and `stream'.  There, the
position angle has positive values and is almost perpendicular to the
observed value.  Finally, we enter the `stream' regime where we define
the elongation of the model as zero. 

For completeness we show the dependency of the position angle
on the initial mass and the Plummer radius in Fig.~\ref{fig:theta}. We
do not see clear trends in this figure but we can see that we have
three regions filled with values and only one of these regions gives
values close to the observed value.  However, all of the points in this
region fail to match the ellipticity of Hercules.  This is demonstrated
in Fig.~\ref{fig:ellippa}. Here we plot ellipticity vs.\ the position
angle and   clearly see the different regimes of Fig.~\ref{fig:sdens}.
While we have a completely bound model we have spherical contours so
the ellipticity is zero and the position angle is not defined (plotted
as zero here as well).  Then we enter the `tidal' regime, where we
see elongated objects, which have the correct position angle (at
least within the errors) but our simulations never reach the strong
ellipticity measured for Hercules.  Then the simulations change into
the `flipped' regime, where the contours of the near field tails are
oriented almost perpendicular to the orbit.  Only in this regime are
we able to match the ellipticity of Hercules, but only at the expense
of matching the position angle.  Finally in the stream
regime we do not have an object and therefore neither position angle
nor ellipticity are defined (both set to zero).  Fig.~\ref{fig:ellippa}
demonstrates that, while it is possible to match either the
ellipticity or position angle, we fail to match the two parameters
simultaneously (see cross symbols, other symbols are discussed in the
following section). 

\section{Best fit model}
\label{sec:gesamt}

\begin{figure}
  \centering
  \epsfig{file=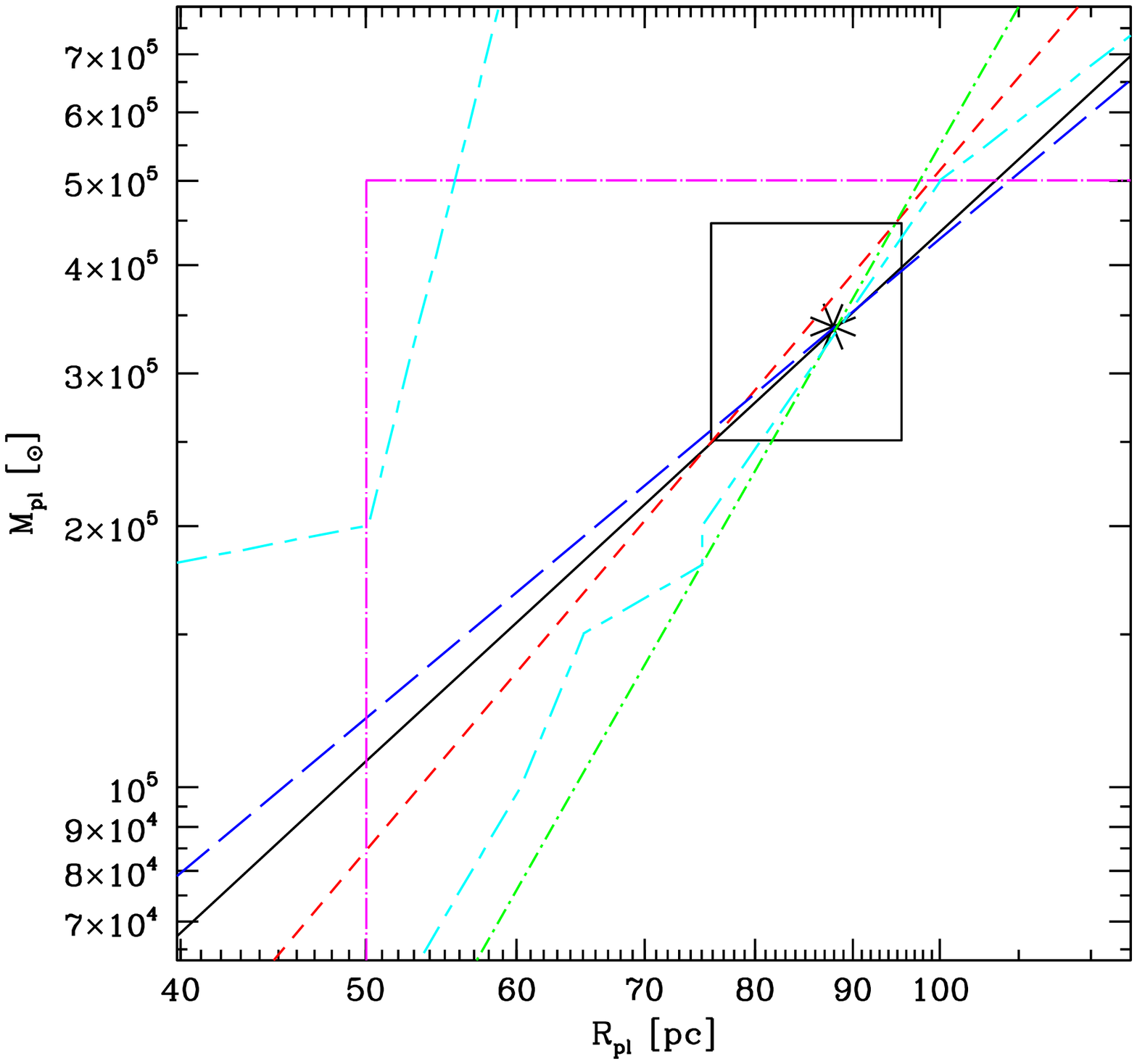, width=8.5cm, height=8.5cm, angle=0}
  \caption{Fitting lines in initial condition space.  The black,
    solid line shows the results matching the final mass, red, short
    dashed line the results matching the central surface brightness,
    blue, long dashed line shows results which match the effective
    radius of Hercules and finally the green, dot-short dashed line
    shows matching solutions for the velocity dispersion.  The magenta 
    dot-long dashed line divides the upper left region which does
    not match the velocity gradient from the lower right, where our
    simulations match the gradient within the errors.  The two cyan
    diagonal short dashed-long dashed lines separate the region, where
    we match the position angle, from the bound region (top left) and
    the stream region (lower right).  Only in between the two lines
    can we match the position angle of Hercules within the errors. The
    black square marks the area in which we expect to find a suitable
    initial model for Hercules.  The black star denotes our best-fit
    model.}  
  \label{fig:gesamt}
\end{figure}

\begin{figure*}
  \centering
  \epsfig{file=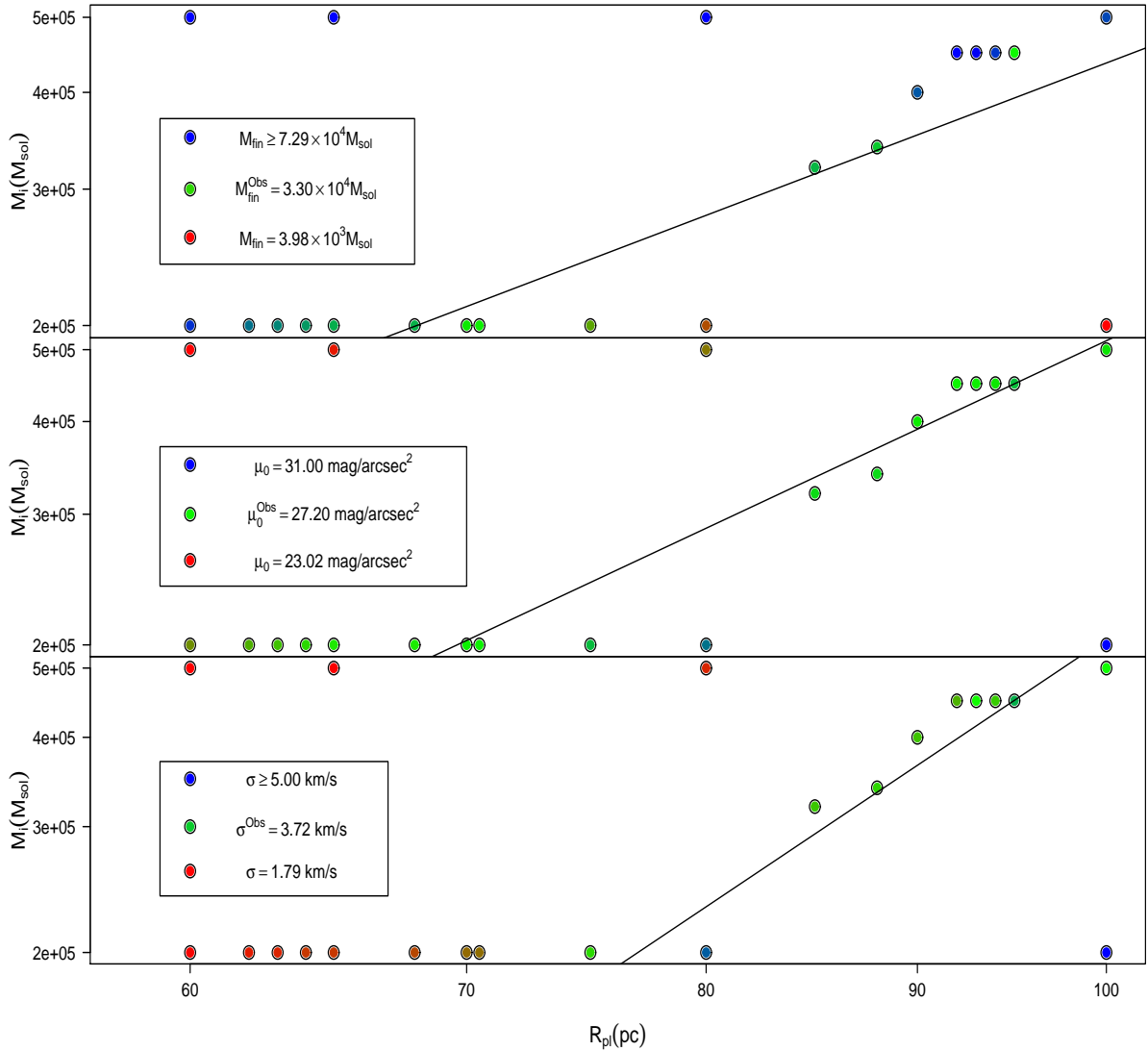, width=15.0cm, height=16.0cm, angle=-90}
  \caption{Plots of inital mass-initial scalelength parameter
    space. Symbol colour relates to final mass (upper panel), central
    surface brightness (middle panel), and velocity dispersion (lower
    panel) as indicated by the key. In each key, the middle row is the
    observed value for Hercules. Black lines join points that should
    produce the observed value, if our fitting technique is
    successful. In practice, simulation points that lie near the black
    line are always close to the observed value, and with increasing
    distance from the line, the simulation values differ increasingly
    from the observed value. This is a strong indication that our
    power-law fitting technique is very effective, and that the
    fitting technique can successfully be used to predict the outcome
    of the simulations.}   
  \label{fig:gesamt2}
\end{figure*}

\begin{figure}
  \centering
  \epsfig{file=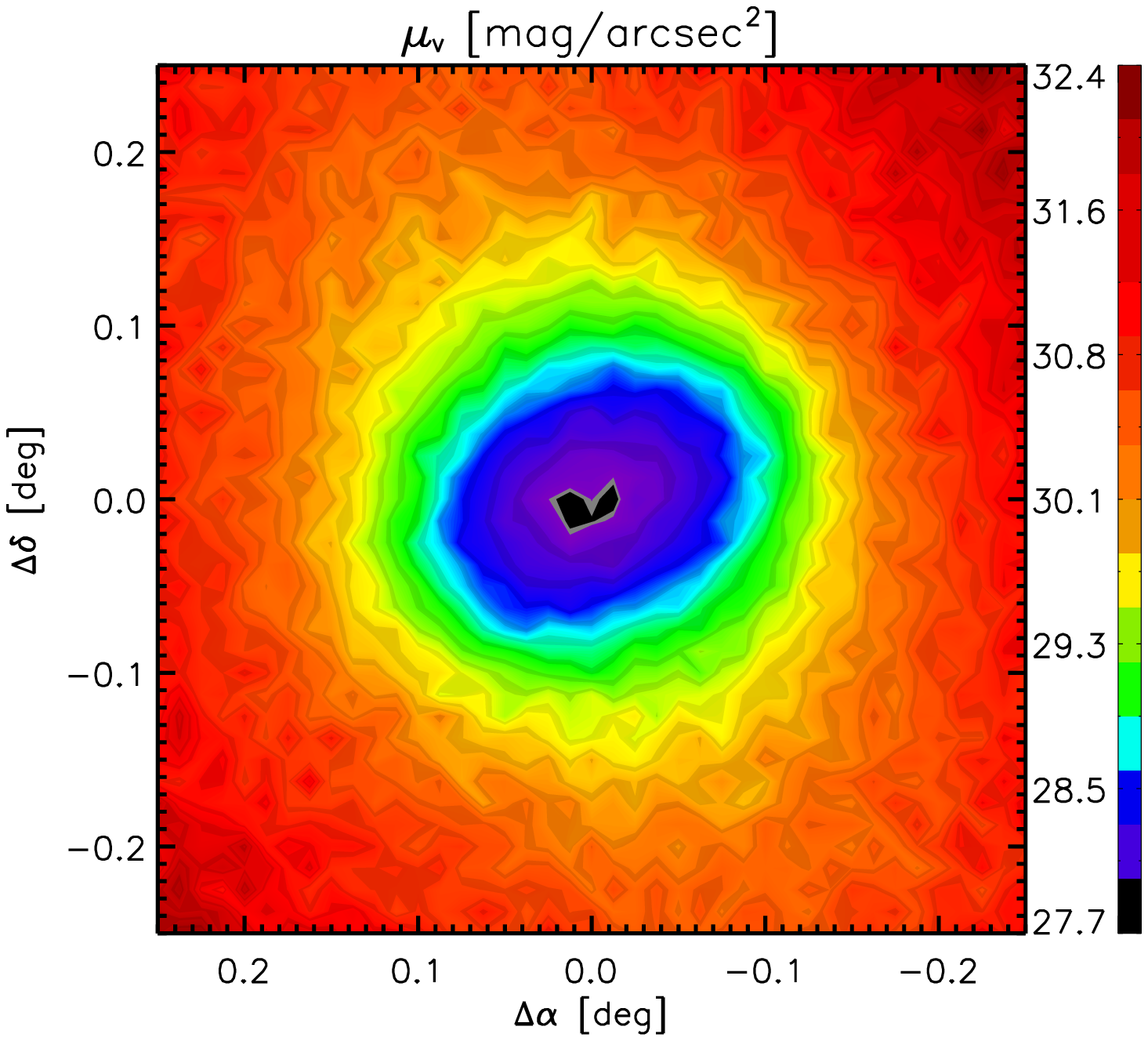, width=7.5cm, height=7.5cm}
  \epsfig{file=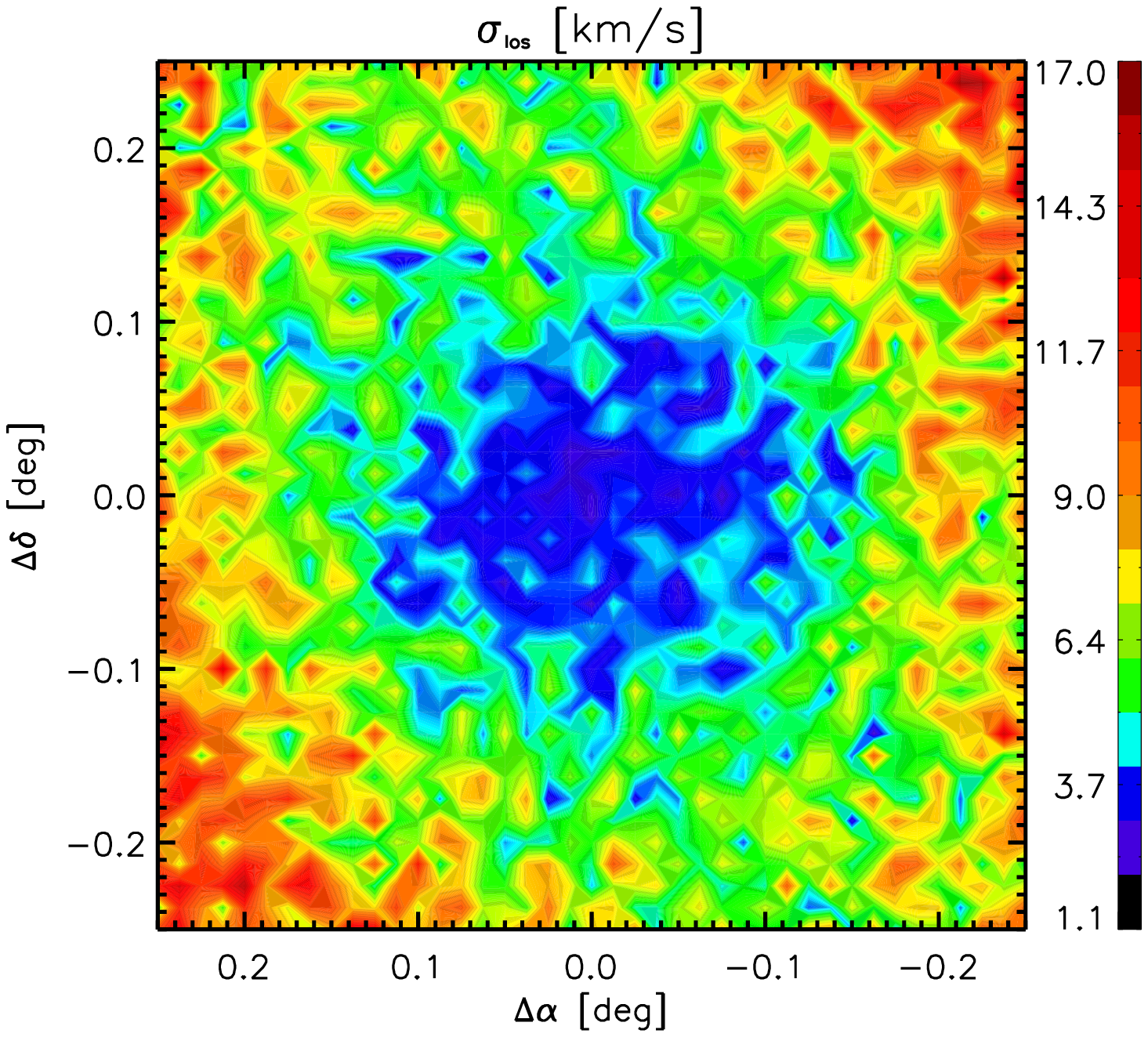, width=7.5cm, height=7.5cm}
  \epsfig{file=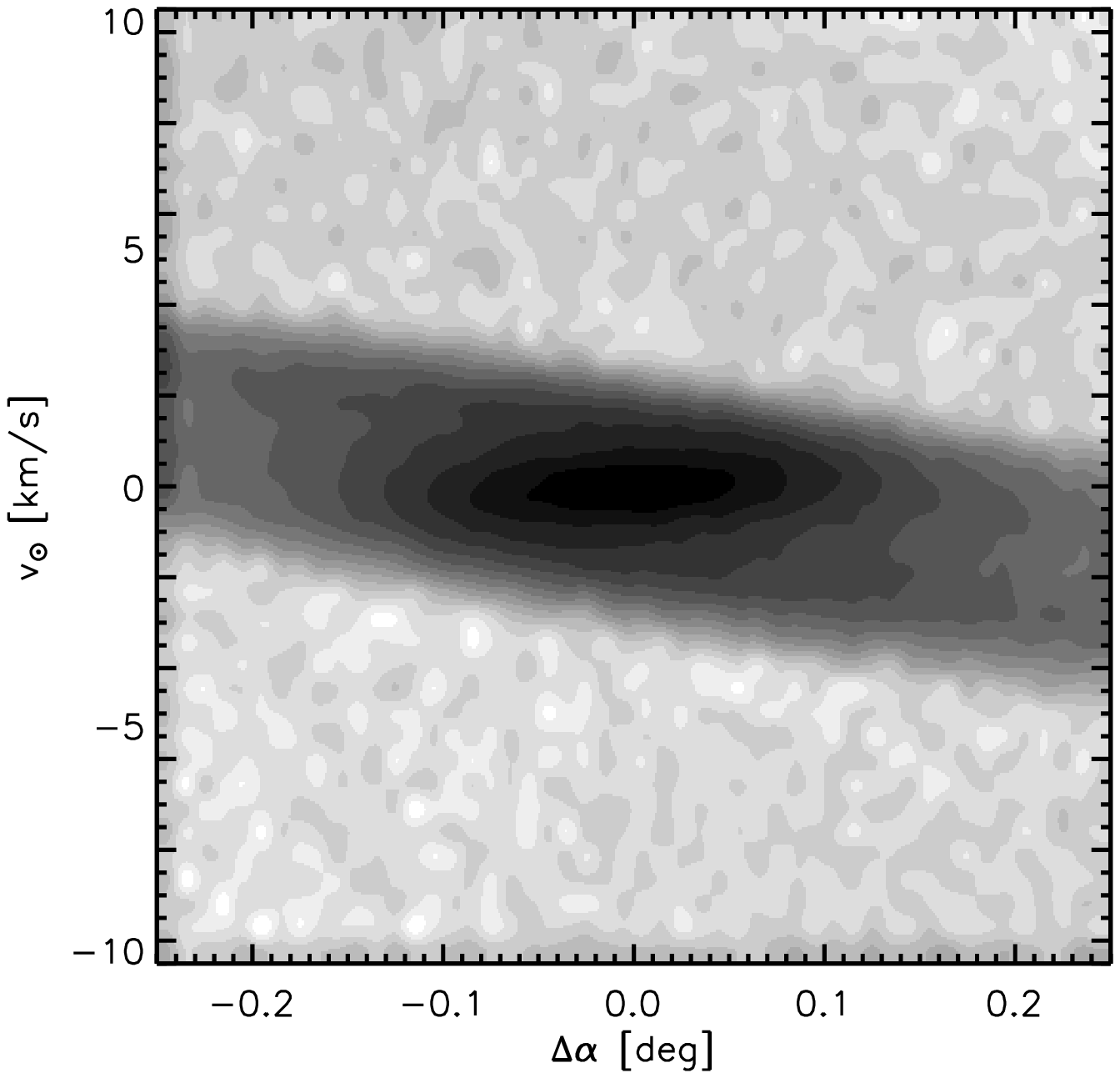, width=7.5cm, height=7.5cm}
  \caption{`Best match model'.  Top: Contour map of the surface
    brightness of our model converting the mass of our particles using
    a pixel resolution of $80$~pixel per degree.  Middle: Contours of
    the two-dimensional distribution of line-of-sight velocity
    dispersions, using the same pixel resolution.  Bottom: Density
    grey-scales of particles in the right ascension -- relative radial
    velocity space.}   
  \label{fig:match}
\end{figure}

Having attempted to fit each observational
parameters separately, we now attempt to combine the results to
produce a best fit model.  For this 
reason we plot all the relations from Eqs.~\ref{eq:mfin}, \ref{eq:mu},
\ref{eq:reff}, and \ref{eq:sigma} as well as the boundaries for
correct solutions regarding position angle and velocity gradient in
one single figure, Fig.~\ref{fig:gesamt}. 

We see that all fitting lines intersect in more or less the same area
of the graph (marked with a black square).  The differences in the
intersection points are all well within the possible errors of the
observational data.  This area is located in the region where we also
match the velocity gradient within its errors and about half of this
area (but with all intersecting points above the division line) falls
into the region where we expect to get the correct position angle.
The correct solution should be found having an initial Plummer 
radius of $76$ to $96$~pc and an initial mass ranging between $2.5
\times 10^{5}$ and $4.4 \times 10^{5}$~M$_{\odot}$ (keeping the
$M/L$-ratio fixed at unity).  
  
We note that the region where the lines are intersecting lies on the
boundary between matching the position angle, and the stream region
where the models flip their position angle through almost a
right-angle (as marked by the lower-right cyan line). Thus the
solution space is always on the brink of flipping 
the position angle, i.e.\ the region of the correct solution is a
solution in which Hercules is already almost destroyed and might 
not survive for much longer.

We present a best fitting model with the initial parameters of $M_{\rm
  pl} = 3.4 \times 10^{5}$~M$_{\odot}$ and $R_{\rm pl} = 88$~pc.  The
final mass of this object, measured in the region described above, is
$4.56 \times 10^{4}$~M$_{\odot}$, which is somewhat higher than the
mass we try to match with our generic $M/L$-ratio.  The central
surface brightness of this object is $27.6$~mag\,arcsec$^{-2}$.
Therefore, the model fits the central surface brightness within the
observational errors.  A two-dimensional contour plot of our object is
shown in the top panel of Fig.~\ref{fig:match}.  The contours are
based on a pixel-map with a resolution of $80$~pixels per degree.  We
also see that the contours show an object which, in the inner parts
is still slightly elongated almost along the orbit, while the outer,
fainter parts have already flipped to the perpendicular direction --
it is an object at the brink of destruction, having lost about
$90$~per cent of its initial mass.  Using the same procedure as for
our other models we arrive at a position angle of only $-63$~degrees
and a small ellipticity of $0.19$.

If we fit a Plummer profile to the surface brightness data calculated
in concentric rings around the centre of the object we get a Plummer
radius of $0.\!^{\rm o}155 \pm 0.\!^{\rm o}007$.  At the distance of
Hercules this translates into a half-light radius of $186 \pm
10$~pc.  This is a rather small value but within $2\sigma$ of the
observed value.  

To demonstrate that our fitting method is effective we perform
  some additional simulations with initial parameters which differ
  from the best fit model. In Fig.~\ref{fig:gesamt2} we demonstrate
  how these choices of initial parameters result in differing final
  mass (upper panel), central surface brightness (middle panel) and
  velocity dispersion (lower panel). Using our fitting procedure, we
  can predict sets of initial mass/initial scalelength that should
  reproduce the observed value. These are shown in each panel as a
  curve. The colour of the symbol shows the actual value measured from
  the simulation - see legend (the middle row of the legend is the
  observed value, and is superscripted by `Obs' to indicate this). In
  all three panels it can be seen that the values measured from the
  simulation for points near the curve agree very well with the
  observed value. Furthermore, with increasing distance from the
  curve, the measured simulation values increasingly differ from the
  observed values. This is strong confirmation that our fitting method
  is effective, and that the results of the fitting can truly be used
  to predict the outcome of the simulation. 
  
In the middle panel of Fig.~\ref{fig:match} we show the contours of
the line-of-sight velocity dispersion.  To achieve this we calculate
the velocity dispersion in each pixel of our 2D figure.  We are able
to do this, as we have more particles in this figure than Hercules
would have stars, as the particles of our simulation represent
equal-mass phase-space elements and not single stars.  We see that in
the central region we have a very low velocity dispersion of about
$1.5$ to $4.0$~km\,s$^{-1}$.  In this region resides the remaining
bound body of our object.  Even though this velocity dispersion is low
compared to the hot regions surrounding that area, it is a very
inflated value.  A bound object with the mass stated above and a
half-light radius of $186 \pm 10$~pc would have a dispersion of about
$0.6$~km\,$s^{-1}$, i.e.\ the velocity dispersion is governed by
unbound stars and their different streaming motions in front, within
and behind the object (as seen by us).  If we compute an overall
velocity dispersion as explained at the beginning of this section we
get a value of $3.14$~km\,s$^{-1}$, which is in excellent agreement
with the observations.  

In the bottom panel of Fig.~\ref{fig:match} we show densities in the
right ascension - radial velocity difference space.  The black area in
the middle shows the remaining bound object, which shows no velocity
gradient, as expected.  But around this area we see the streaming
motion of the orbit very clearly as a dark grey area.  A parallel line
through the middle of this area would lead to a velocity gradient of
about $-8$~km\,s$^{-1}$\,degree$^{-1}$.  Using the calculation we
describe at the beginning of this section, we still obtain a value of
$-5.7$~km\,s$^{-1}$\,kpc$^{-1}$, which is within the errors stated by
\citet{jin10}. 

In summary, our best-fit model can well match the observed values of
the luminosity, central surface brightness, effective radius, velocity
dispersion and velocity gradient.  This demonstrates that the
technique for finding a best-match model, as described in
Sec.~\ref{sec:technique}, is effective at providing the means to
simultaneously match multiple properties in a systematic
manner.  However, despite this success, the technique was unable to
simultaneously match the position angle and ellipticity.  It is not
impossible that one cause of this failure might be that the orbit we
have assumed is not that of the real Hercules.  However, to support
such a hypothesis, we must first demonstrate that the
ellipticity-position angle failure continues to occur if we
significantly change the duration of the simulation, or if we
replace our cored progenitor model with a more cuspy profile. 

\subsection{Infall 5 Gyr ago}
\label{sec:5gyr}

\begin{figure}
  \centering
   \epsfig{file=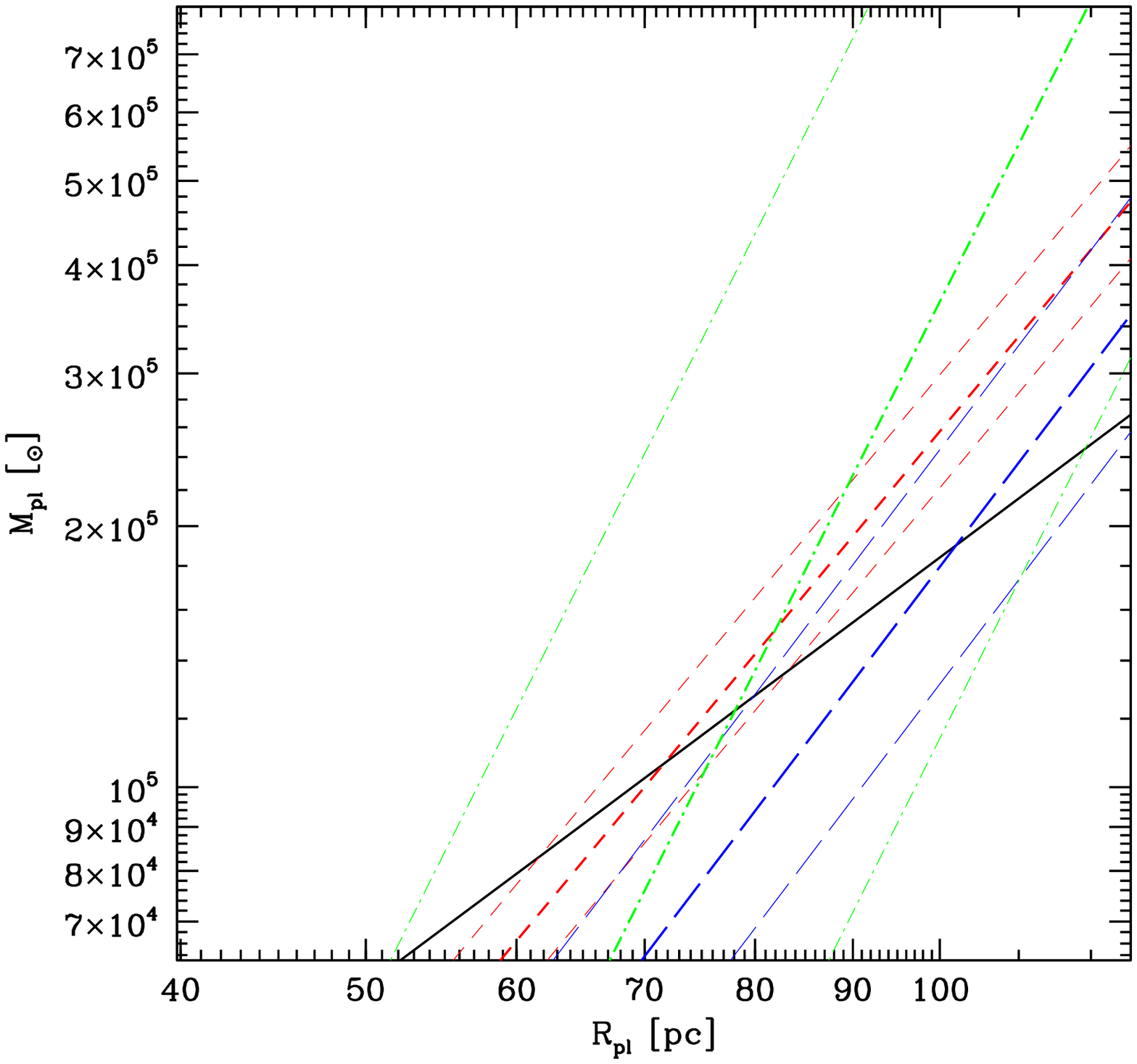, width=8.5cm, height=8.5cm}
  \caption{Lines of matching initial models for an orbital time of
    $5$~Gyr.  Color and line-style code as in Fig.~\ref{fig:gesamt}.
    Additionally we plot the lines if we shift the power-laws by one
    sigma in their zero-point.  We see that again we find a region
    where all lines intersect.  Here we expect to find a model that
    matches these particular observables simultaneously.}
  \label{fig:5gyr}
\end{figure}

The true infall time of Hercules into the Milky Way is rather
uncertain.  So far we have only considered an infall 10~Gyr ago.  By
considering a more recent infall time of 5~Gyr, we can test if our
technique for finding a best match model is robust to varying the
duration of the simulation.  It is not suprising that the initial
parameters of our new best match model will be different than in the
10~Gyr case - in order to match today's properties of Hercules, the
new model must tidally evolve in half the time period.  It also allows
us to see, if the ellipticity-position angle failure that our best fit
model suffers continues when we change the mass loss rate quite
considerably. 

We perform 27 simulations starting at
$t=-5$~Gyr.  We repeat the same technique as described previously but
with the new simulations.  First we derive power-laws of matching
initial conditions for the mass: 
\begin{eqnarray}
  \label{eq:m2}
  M_{\rm pl} & = & 95^{+12}_{-11} \cdot R_{\rm pl}^{1.64 \pm 0.03},
\end{eqnarray}
for the central surface brightness:
\begin{eqnarray}
  \label{eq:u2}
  M_{\rm pl} & = & 1.31^{+0.21}_{-0.19} \cdot R_{\rm pl}^{2.65 \pm 0.04},
\end{eqnarray}
the effective radius:
\begin{eqnarray}
  \label{eq:r2}
  M_{\rm pl} & = & 0.28^{+0.10}_{-0.08} \cdot R_{\rm pl}^{2.91 \pm 0.07},
\end{eqnarray}
and the velocity dispersion:
\begin{eqnarray}
  \label{eq:v2}
  M_{\rm pl} & = & 0.61^{+1.32}_{-0.42} 10^{-3} \cdot R_{\rm pl}^{4.4 \pm 0.3}.
\end{eqnarray}
The errors on these fitting lines are quite large as they are based on 
much less simulations than our main study.  Especially, the
zero-points of the power-laws are quite ill determined.  For this
reason we plot in Fig.~\ref{fig:5gyr} the fitting lines plus the lines
we would obtain if we add or subtract one-sigma in the zero-point.  

Then we combine the power laws to find the best-fit model as shown in
Fig.~\ref{fig:5gyr}.  We find we are able to find a suitable match,
despite the large change in simulation time.  However, this time we
match only four of the main parameters -- namely final mass, central
surface brightness, effective radius, and velocity dispersion. 

As in the main study we see that all four lines intersect (within
their errors) in the same region of the diagram.  We, therefore,
infer that should the orbital time of Hercules have been only $5$~Gyr
we would have to search the correct matching model in a region where
its initial parameters are about $R_{\rm pl} \approx 70-80$~pc and
$M_{\rm pl} \approx 1.0-1.5 \cdot 10^{5}$~M$_{\odot}$.

Comparing this with our previous findings shows that we do not need to
change the initial scale-length dramatically but we have to search at
much lower initial masses (factor $2-4$).  By having a lower mass, the
model is much more easily destroyed by the tides of the MW, and
therefore reaches the same point of destruction we want to observe in
a much shorter orbital time.

Unlike in the best-fit model, we cannot match the velocity gradient.
It is consistently weaker than in the 10~Gyr duration simulations
(although still within the published errors).  The position angle and 
ellipticity show the same behaviour as seen in the $10$~Gyr
simulations:  with the shorter simulation time we still cannot match
the position angle and the ellipticity at the same time.  Simulations
with the correct ellipticity show the `flipped' orientation and vice
versa (e.g. see red triangles in Fig.~\ref{fig:ellippa}). 

\subsection{Hernquist models}
\label{sec:hern}

\begin{figure}
  \centering
  \epsfig{file=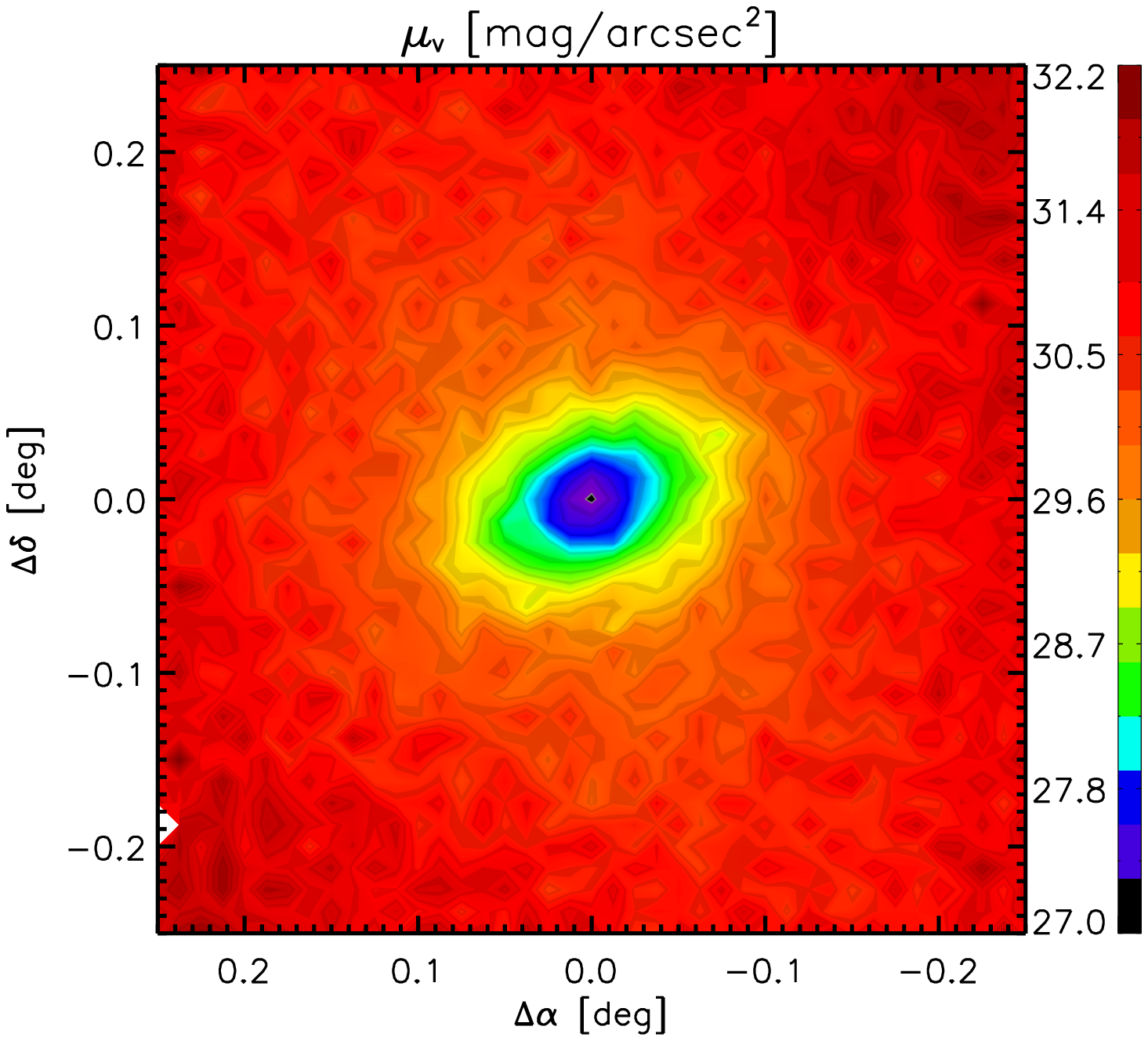, width=8.5cm, height=8.5cm}
  \caption{Contour plot of the surface brightness of the best-match
    Hernquist model -- as in the upper panel of Fig. \ref{fig:match}
    except here for the Hernquist model.} 
  \label{Hernmodel}
\end{figure}

We furthermore tried to match the observables using a
cuspy initial model.  For a cuspy model, we used Hernquist spheres
instead of Plummer models.  We use an orbital time of $10$~Gyr to
compare them with our main results. 

By trial and error, we vary the initial mass and initial Hernquist
scalelength, to find a model that closely matches the observed central
surface brightness and luminosity of Hercules.  Our `best-match'
Hernquist model is shown in Fig.~\ref{Hernmodel}.  This model has a
similar final appearance to the best fit Plummer model, although its
surface brightness falls off more quickly (e.g.\ compare with the upper
panel of Fig. \ref{fig:match}).  The best-match Hernquist model has a
final mass M$_{\rm{fin}}=2.9\times10^4~$M$_\odot$, and central surface
brightness $\mu_0=27.0$~mag~arcsec$^{-2}$ -- each within one-sigma of
the observed values. 

Furthermore, the velocity gradient is also a reasonable match ($\Delta 
v_{\rm{r}} = -6.9$~km~s$^{-1}$~kpc$^{-1}$).  However, the velocity
dispersion $\sigma_{\rm{los}} = 7.4$~km~s$^{-1}$ -- over a factor of two
{\it{too high!}}.  But, looking at the actual distribution of the
velocities we see that in the Hernquist models these high dispersions
are caused by a few stars with large differences in velocities (in
contrary to the Plummer models).  If we adopt a clipping function for
those velocity tails we obtain a distribution with a FWHM (full width
half maximum) of about
$3$~km\,s$^{-1}$, i.e.\ a dispersion which matches the observed one.  

The initial mass and scalelength of our best-match
Hernquist model is M$_{\rm{init}} = 8.0 \times 10^{5}$~M$_\odot$, and
R$_{\rm{hern}} = 0.175$~kpc respectively.  Thus, the progenitor Hernquist
model is over a factor of two more massive than the progenitor Plummer
model.  Therefore in order to match the final observed mass, the
Hernquist model must lose more than a factor of two more mass than the
Plummer model, and this could explain the enhanced velocity dispersion
-- the Hernquist model is closer to being completely disrupted. 

The green circular symbols in Fig. \ref{fig:ellippa} are the results
of the various Hernquist models we conducted in the course of finding
our best-fit model.  Our best-match Hernquist model is the symbol with
position angle $\theta = -67^\circ$, and an ellipticity
$\epsilon = 0.27$.  Clearly the more cuspy Hernquist models suffer the
same fate as the more cored Plummer models -- a failure to match the
observed ellipticity unless the model is so unbound that it has
flipped in position angle.  Indeed the low surface brightness, outer
contours surrounding our best-match Hernquist model show the same
flipped shape as was seen in the best fit Plummer model (e.g. again,
compare Fig.~\ref{Hernmodel} with the upper panel of
Fig.~\ref{fig:match}). 

In summary, we find that the problem of simultaneously matching the
positional angle and ellipticity (e.g.\ see
Fig.~\ref{fig:ellippa}) is robust to significant changes in the infall
time (we change it by a factor of 2), and also to varying the
cuspiness of the initial progenitor model.  We also see flipped
outer-most contours, tracing the unbound stars, in all three best
match models (10~Gyr, 5~Gyr, and the Hernquist model).  We believe
that the flipping of the unbound streams is actually a property of the
orbit on which we assume the models follow.  If so, this indicates it
may be impossible to solve the ellipticity-position angle problem,
assuming an initially spherical progenitor that is tidally stripped
{\it{while moving along the published orbit\footnote{For the published
      orbit, we use the most probable orbital parameters given in
      \citet{jin10}, and we do not consider the errors.}.}}  We
elaborate on the 
possibility that the orbit is incorrect in the following section.  

\section{Discussion and Conclusions}
\label{sec:conc}

We have used a new method to find a suitable model to try
to reproduce the observables of the dwarf galaxy Hercules.  Instead
of trial and error around a possible solution, we made use of a wide 
parameter space of initial conditions and analysed the general
behaviour of every observable as a function of the initial
parameters to assess their power-law dependencies in the region of
interest.  We have shown that we find a relatively small region
(smaller than the observational errors) in initial parameter space,
where we match several of the observables simultaneously.

We have shown that {\it{for the given orbit}}, the new technique
is successful in finding a best-match model for Hercules.  We
  emphasise that the orbit we consider is based on the most probable
  orbit given in \citet{jin10}, and we do not consider possible
  alternative orbits within the error bounds of the orbital
  parameters.  {\bf We also do not take into account that the orbit
    may have been different in the past (e.g.\ got changed by a close 
  encounter with another dwarf satellite halo).}  In our 
case, Hercules has an initial mass of $3.4 \times
10^{5}$~M$_{\odot}$ and an initial scale-length of $88$~pc to match
the observables today.  These values will only change slightly if
the orbit turns out to be slightly different \citep[see e.g.\ the
change in initial mass in the models of][]{fell07b} or the orbiting
time is only slightly different.  We find that, unsuprisingly, 
shortening the orbital time causes a change in the initial mass and 
scalelength required to match the luminosity of Hercules. In our case 
halving the orbital time led to an initial model with a factor 2-3
times lower initial mass, and a slightly smaller scalelength
($\sim 80$~pc). 

The best-match model is successful in matching a large number of the 
observed parameters of Hercules, including luminosity, central surface 
brightness, effective radius, velocity dispersion, and velocity gradient.
This clearly demonstrates the power of the new systematic technique used
to find the best-match model.  However, despite the thoroughness of the 
technique, we find it impossible to match the observed ellipticity and 
position-angle simultaneously in any of our models. 

All models on the published orbit show a similar behaviour.  While a
bound core remains after 10~Gyr, the position angle is close to the 
observed value, but the core is too round resulting in an ellipticity
which is much lower than the observed value.  However, models that are 
slightly more tidally disrupted after 10~Gyr can match the observed
ellipticity, but in the process their position angle flips to almost
perpendicular to the observed value.  This behaviour seems to be an
inherent property of the orbit of the galaxy.  In fact the
ellipticity-position angle problem persists, assuming the published
orbit, even when we consider an infall time half as long, and even
when we exchange the progenitor model for a much more cuspy density
profile.  We note that by changing the infall time, we have altered
the mass loss rate considerably, and this could be considered broadly  
equivalent to including other mass loss mechanisms, such as two body
encounters, or allowing for a Milky Way potential that evolves with
time.  {\bf We have not included the possibility, that Hercules may
  have changed its orbit in the past due to an encounter with another
  satellite halo.  In this case the previous orbit would be completely
  unknown to us. But, again, we do not believe that such a scenario
  would alter our conclusions.  In fact, this scenario would have
  elongated Hercules in some random orientation before the encounter
  and the observed ellipticity today would even be harder to match.

  Despite changing the duration of the simulations by a factor of 2,
  and significantly changing the rate of mass loss, the final models
  suffer the same issues with simultaneously matching the ellipticity
  and position angle.  However, in both the 10 Gyr and 5 Gyr
  simulations, the models end up at the same position within the
  potential of the Milky Way.  This likely suggests that the problems
  with matching the ellipticity and position angle occur due to the
  shape of the potential field in which the models sit now, and not
  due to their earlier tidal history. } 

We believe that the flipping of the position angle, to beyond the
observed value, which is seen in the unbound stars of all our models,
is actually a property of the chosen orbit.  If so, it may be
impossible to solve the ellipticity-position angle problem while using 
the published orbit -- even considering other spherical progenitor
models.  This naturally leads us to one of three scenarios: 
\begin{enumerate}
\item The Hercules progenitor has an intrinsically flattened stellar
  distribution that is shielded from tidal distortion by a massive
  dark matter halo.  In this case, the elongation of the stars
  provides us with no information on the orbit and the published
  orbit, which is based on the assumption of tidal distortion, is
  almost certainly incorrect. 
\item The Hercules progenitor has an intrinsically flattened stellar
  distribution, but no dark matter.  This would require that the
  intrinsic elongation is near perfectly aligned with the elongation
  from tidal features, and therefore we consider this case to be
  highly unlikely. 
\item The Hercules progenitor was spherical and dark matter free, as
  we considered in this study.  The progenitor was later elongated by
  tidal disruption along the published orbit.  However, in this case
  our models cannot match the observed ellipticity.  If the true
  ellipticity of Hercules is lower 
  than the quoted values in \cite{col07}, then this third scenario is
  possible.  
\end{enumerate}
In fact all three scenarios lead to the same conclusion: if Hercules
is truly flattened to the extent observed, it is highly likely that
the orbit we assume throughout this paper is incorrect.\\  

\noindent
{\bf Acknowledgments:}
MF acknowledges financial support of FONDECYT grant no.\ 1095092,
1130521 and BASAL PFB-06/2007.  RS acknowledges financial support of
FONDECYT grant no.\ 3120135.  GC acknowledges financial support of
FONDECYT grant no.\ 3130480. RC acknowledges financial support through 
an ESO Comite Mixto grant. 

\bibliographystyle{mnras}

\label{lastpage}

\end{document}